\documentclass[lettersize,journal]{IEEEtran}

\usepackage{amsmath, graphicx}
\usepackage{amsfonts, amsthm, amssymb, mathrsfs, mathtools}
\usepackage{bm}
\usepackage{algorithm, algorithmic}
\usepackage{setspace}
\usepackage{cite}
\usepackage{subfig}
\usepackage{array}
\usepackage{booktabs}
\usepackage{subfiles}
\usepackage{multirow}
\usepackage{cases}
\usepackage{makecell}
\usepackage{bbding}
\usepackage{footmisc}
\usepackage{rotating}
\usepackage{caption}
\usepackage{ulem}
\usepackage{diagbox}
\usepackage{xcolor}
\usepackage{tikz}
\usepackage{utfsym}

\usepackage[english]{babel}
\usepackage{url}
\usepackage[colorlinks,
linkcolor=magenta,
anchorcolor=magenta,
citecolor=magenta,
]{hyperref}

\newtheorem{theorem}{Theorem}

\newcommand{\X}{\mathcal{X}}
\newcommand{\Y}{\mathcal{Y}}
\newcommand{\Z}{\mathcal{Z}}
\newcommand{\la}{\mathcal{L}}
\newcommand{\G}{\mathcal{G}}
\newcommand{\R}{\mathbb{R}}
\newcommand{\brm}[1]{\bm{\mathrm{#1}}}
\newcommand{\Dn}{\downarrow}
\newcolumntype{M}[1]{>{\centering\arraybackslash}m{#1}}
\newcolumntype{L}[1]{>{\raggedright\arraybackslash}m{#1}}
\newcommand{\tb}[1]{\textbf{#1}}
\newcommand{\ul}[1]{\underline{#1}}

\newcommand{\dui}{\usym{1F5F8}}
\newcommand{\bdui}{\usym{2717}}

\begin{document}

\title{Variational Zero-shot Multispectral Pansharpening}

\author{Xiangyu Rui, Xiangyong Cao, Yining Li, and Deyu Meng.
	\thanks{This work was supported in part by the National Key R\&D Program of China (2022YFA1004100) and in part by China NSFC Projects under Contract 12226004 and 62272375. (\textit{Corresponding author: Xiangyong Cao; Deyu Meng.})}
	\thanks{Xiangyu Rui and Yining Li are with the School of Mathematics and Statistics and Ministry of Education Key Lab of Intelligent Networks and Network Security, Xi'an Jiaotong University, Xi'an 710049, Shaanxi, China. (email: xyrui.aca@gmail.com, yiningli666@gmail.com)}
	\thanks{Xiangyong Cao is with the School of Computer Science and Technology and Ministry of Education Key Lab For Intelligent Networks and Network Security, Xi'an Jiaotong University, Xi'an 710049, Shaanxi, China. (email: caoxiangyong@xjtu.edu.cn)}
	\thanks{Deyu Meng is with the School of Mathematics and Statistics and Ministry of  Education Key Lab of Intelligent Networks and Network Security, Xi’an Jiaotong University,  Xi’an 710049, Shaanxi, China, and Macao Institute of Systems Engineering, Macau University of Science and Technology, Taipa, Macao. (email: dymeng@mail.xjtu.edu.cn)}
	}	

\markboth{Journal of \LaTeX\ Class Files,~Vol.~14, No.~8, August~2024}%
{Shell \MakeLowercase{\textit{et al.}}: A Sample Article Using IEEEtran.cls for IEEE Journals}


\maketitle

\begin{abstract}
	Pansharpening aims to generate a high spatial resolution multispectral image (HRMS) by fusing a low spatial resolution multispectral image (LRMS) and a panchromatic image (PAN). 
	The most challenging issue for this task is that only the to-be-fused LRMS and PAN are available, and the existing deep learning-based methods are unsuitable since they rely on many training pairs. Traditional variational optimization (VO) based methods are well-suited for addressing such a problem. They focus on carefully designing explicit fusion rules as well as regularizations for an optimization problem, which are based on the researcher's discovery of the image relationships and image structures. Unlike previous VO-based methods, in this work, we explore such complex relationships by a parameterized term rather than a manually designed one. Specifically, we propose a zero-shot pansharpening method by introducing a neural network into the optimization objective. This network estimates a representation component of HRMS, which mainly describes the relationship between HRMS and PAN. In this way, the network achieves a similar goal to the so-called deep image prior because it implicitly regulates the relationship between the HRMS and PAN images through its inherent structure. We directly minimize this optimization objective via network parameters and the expected HRMS image through alternating minimization. Extensive experiments on various benchmark datasets demonstrate that our proposed method can achieve better performance compared with other state-of-the-art methods. The codes are available at \url{https://github.com/xyrui/PSDip}.
\end{abstract}


\section{Introduction}
\IEEEPARstart{M}{ultispectral} image is a kind of optical image captured by special sensors. It records the wavelengths from specific electromagnetic spectrums. Some wavelengths, e.g., thermal infrared, can not be detected by human eyes as visible lights. Thus, multispectral images can provide more information about the objects than general RGB images and they have been applied in many fields, including medicine \cite{application-medicine}, agriculture \cite{application-arg1, application-arg2}, food industry \cite{application-food}, etc.

\begin{figure}[t]
	\centering
	\includegraphics[width=8cm]{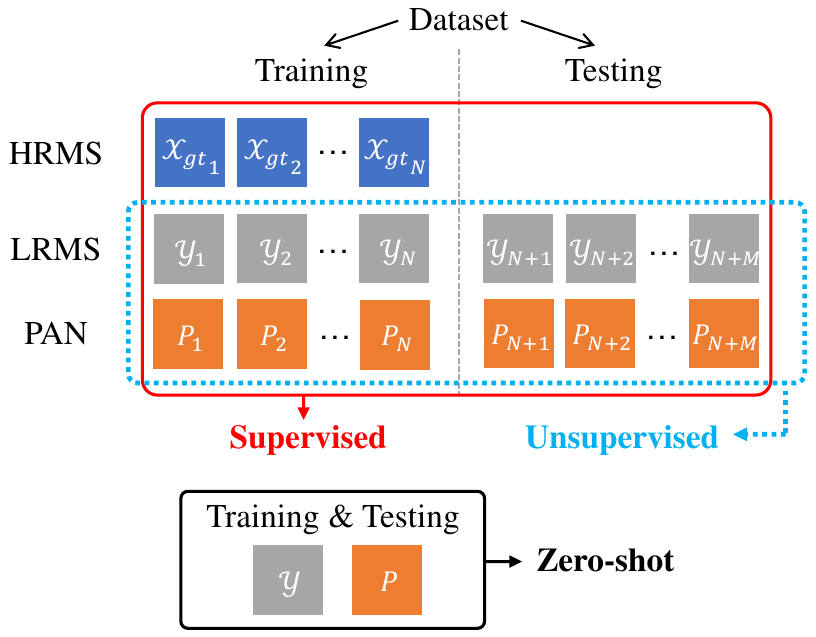}
	\caption{Comparison between supervised, unsupervised and zero-shot multispectral pansharpening methods from the perspective of dataset.}
	\label{fig-sup-un-zs}
\end{figure}

Multispectral pansharpening aims to obtain a high spatial resolution multispectral image (HRMS) by fusing a low spatial resolution multispectral image (LRMS) and a panchromatic image (PAN). This problem comes with the fact that acquiring HRMS data typically involves higher sensor capabilities, longer time requirements, and increased expenses. In contrast, the LRMS covers the same wavelength range as HRMS but has a lower spatial resolution, and the PAN has the same high spatial resolution as HRMS but is in a single band. Since LRMS and PAN are easier to obtain, technically fusion of LRMS and PAN is a cost-effective way to obtain HRMS, which has constantly attracted the attention of researchers.

The multispectral pansharpening methods are mainly divided into four classes \cite{Review}. Among them, the component substitution (CS) based methods, the multiresolution analysis (MRA) based methods and the variational optimization (VO) based methods are also called traditional methods in this work. They do not contain learnable parameters and thus also do not require any training data. Deep learning (DL) based methods require large number of HRMS/LRMS/PAN training pairs to learn the neural network that generates HRMS. Unsupervised DL-based methods eliminate the reliance on HRMS. Instead, they use sufficient number of LRMS/PAN training pairs to learn the underlying relationships between HRMS, LRMS, and PAN of a specified scene \cite{Deep-image-interpolation-A-unified-unsupervised-framework-for-pansharpening, PANGAN, LDPNet, PercepPan, PGMAN,Unsupervised-MunGAN, UPanGAN}. The training pairs are comprised of part of this scene. The rest of the part is used to evaluate the trained network. When facing a new scene, the network should be retrained using LRMS/PAN pairs from this scene. More recently, zero-shot methods \cite{ZSPan}\cite{Zero-Sharpen} have been researched. They also leverage neural network to learn the data relationship. However, they handle the most fundamental yet also challenging situation where only a single pair of LRMS and PAN is available for both training and testing. In Fig. \ref{fig-sup-un-zs} we illustrate the difference between supervised, unsupervised and zero-shot methods from the perspective of training and testing dataset.

This work handles the zero-shot situation that no other data except the observed LRMS and PAN is available. Following the VO-based works, we focus on modeling the HRMS/PAN relationship. Early VO-based works simply assume linear form in the spectral dimension. The assumption soon did not meet the higher expectation of fusion performance because 1) the linear assumption may not reflect the actual situation and 2) there still leaves much more image information that the linear assumption can not effectively provide. Thus, a series of methods with further exploration were then developed~\cite{A-variational-approach-for-pan-sharpening, A-new-pansharpening-method-based-on-spatial-and-spectral-sparsity-priors, Pan-sharpening-with-a-hyper-Laplacian-penalty, A-variational-pan-sharpening-method-based-on-spatial-fractional-order-geometry-and-spectral-spatial-low-rank-priors, A-variational-pan-sharpening-with-local-gradient-constraints, High-quality-Bayesian-pansharpening, SIRF-Simultaneous-satellite-image-registration-and-fusion-in-a-unified-framework, A-new-variational-approach-based-on-proximal-deep-injection-and-gradient-intensity-similarity-for-spatio-spectral-image-fusion, VO+Net, LRTCFPan, SFNLR}, e.g., the HRMS/PAN relationship is considered on the gradient field. Recently, inspired by CS- and MRA-based methods, HRMS has also been represented using the hadmard product of a coefficient tensor and a transformed PAN that has the same size as HRMS~\cite{SFNLR}. The biggest advantage of this representation is that, unlike linear form, it theoretically could have zero approximation error for any HRMS/PAN pairs. The problem is that if the coefficient tensor is a completely free component, such representation, however, is noninformative and does not provide effective image information about HRMS at all. In previous traditional works, the coefficient tensor is usually pre-estimated by different ingenious methods. The model performance largely relies on the setting of coefficient tensor. However, the pre-estimation process is usually independent from the optimization problem, which would limit the model performance.

To address the above problem, we propose a new zero-shot multispectral pansharpening method which is based on a new proposed VO-based optimization framework. Specifically, we also represent HRMS by the product of a coefficient and a transformed PAN. But we place a regularization term about the coefficient in the general optimization objective for pansharpening. Then, the coefficient together with the expected HRMS are both treated as the variables to be optimized. Such formulation achieves two goals. First, it delivers effective information from PAN to HRMS because the coefficient is constrained. Second, the structure of HRMS is also indirectly regularized via the regularization on the coefficient. Actually, we proof that without changing the restored HRMS in the minimum point of the optimization problem, the regularization on HRMS can be completely absorbed into the regularization on the coefficient. Then, based on the recent discovery that the coefficient contains image-looking structures~\cite{SFNLR}, we consider the DIP-type regularization \cite{DIP}. Specifically, the coefficient tensor is predicted by a neural network whose structures can naturally deliver the image prior. In this way, the network parameters and the expected HRMS are optimized by minimizing the optimization objective. We simply iteratively update network parameters and the expected HRMS using gradient descent-based methods. Since the proposed method contains neural network that needs to be trained, it is a zero-shot method. Compared with existing zero-shot methods, the proposed method has more concise and compact form.

In conclusion, the contributions of this work are listed as follows.
\begin{itemize}
	\item A zero-shot multispectral pansharpening method is proposed, which does not require additional training data other than the observed LRMS and PAN. Specifically, we formulate a new optimization objective in which the regularization of the coefficient is contained. The coefficient and the expected HRMS are taken as variables to be optimized within the one optimization problem. 
	\item A DIP-type regularization is considered for the coefficient in the formulated optimization objective. Specifically, we use a neural network to predict the coefficient and the network structure itself could provide necessary prior to reconstruct the coefficient. Then, the network parameters are optimized together with the expected HRMS.
	\item An easy-to-implement strategy, i.e., alternating minimization, is used to solve the formulated optimization problem. To stabilize the optimization process, the network is first initialized before the alternating minimization.  
	\item Experiments on several benchmark datasets show that the proposed method is convenient for handling diverse datasets and achieves comparable or superior performance to existing state-of-the-art (SOTA) methods.  
\end{itemize}

The organization of this work is as follows. In Sec. \ref{sec-Related works}, we review the previous pansharpening methods and DIP-related methods. In Sec. \ref{sec-Main-method}, the pansharpening problem is introduced and the proposed method is analyzed. In Sec. \ref{sec-experiments}, experiments on four datasets are conducted to verify the effectiveness of the proposed method. Besides, we also analyze the parameter setting and conduct several ablation studies. In Sec. \ref{sec-conclusion}, we summarize the proposed method.

\section{Related Works}\label{sec-Related works}
\subsection{Multispectral pansharpening}
Multispectral pansharpening mainly contains four class, i.e., CS-based methods, MRA-based methods, VO-based methods and DL-based methods. The specific form of CS-based methods and MRA-based methods can be mainly described as the sum of an upsampled LRMS and an injection component \cite{A-critical-comparison-among-pansharpening-algorithms}. For CS-based methods, according to how the injection gains and the intensity components are calculated, it has given rise to various classic methods, including Brovey transform \cite{Brovey}, Gram-Schmidt \cite{GS} and its variation GSA \cite{GSA}, BDSD \cite{BDSD}, PRACS \cite{PRACS}, etc. Recently, \cite{Robust-band-dependent-spatial-detail-approaches-for-panchromatic-sharpening} improves BDSD to handle the issue where HRMS has more spectral bands. MRA-based methods consider injecting spatial details extracted from the difference of PAN and its low-pass filtered version into the upsampled LRMS. Varying from injection gains and low-pass filter, typically MRA-based methods include HPF \cite{HPF}, Indusion \cite{Indusion}, GLP \cite{GLP}, SFIM \cite{SFIM}, etc. 

VO-based methods take pansharpening as an inverse problem. In this way, an optimization problem is built. It usually contains data fidelity terms and regularization terms. The spectral fidelity term describes the degenerative process from HRMS to LRMS. The spatial fidelity term describes the relationship between HRMS and PAN. Compared with spectral fidelity, the spatial fidelity term is more challenging to design. Previously, the first VO-based method proposed in \cite{A-variational-model-for-P+XS-image-fusion} assumed that PAN is the linear combination of the spectral channels of HRMS. This assumption does not imply that it exactly matches the real situation, but it can be said to be a good approximation and easy to implement. Thus, the idea has been utilized in many VO-based methods. \cite{A-new-pan-sharpening-method-using-a-compressed-sensing-technique} takes this linear assumption into a compressed sensing model in which the observations are represented by a dictionary and a sparse coefficient. \cite{A-new-pansharpening-algorithm-based-on-total-variation} combines this assumption with total variation regularization on the HRMS and formulates an optimization problem. \cite{A-variational-pansharpening-approach-based-on-reproducible-kernel-Hilbert-space-and-heaviside-function} extends the idea of continuous modelling proposed in \cite{Single-image-super-resolution-via-an-iterative-reproducing-kernel-Hilbert-space-method} to pansharpening by representing the HRMS in the reproducible kernel Hilbert space. In \cite{Multiband-remote-sensing-image-pansharpening-based-on-dual-injection-model}, the linear assumption is combined with a dual-injection model that extracts high-frequency components from the PAN.

After all, the direct linear modelling between HRMS and PAN has its limitations. \cite{A-variational-approach-for-pan-sharpening} proposes to transfer the linear assumption to the gradient field, which is a generalized form of the linear assumption on the original image field. \cite{A-new-pansharpening-method-based-on-spatial-and-spectral-sparsity-priors} adds the extended total variation constraint on HRMS and PAN except for the linear one, which aligns the high frequency of them. \cite{Pan-sharpening-with-a-hyper-Laplacian-penalty} uses $\ell_{1/2}$ norm to measure the difference of the gradients on the linear combination of HRMS and the PAN. The gradients also include two diagonal directions besides the usual horizontal and vertical directions. \cite{SIRF-Simultaneous-satellite-image-registration-and-fusion-in-a-unified-framework} proposes group sparsity constrain on the difference of gradients in HRMS and the extended PAN to encourage the so-called dynamic gradient sparsity. \cite{A-variational-pan-sharpening-method-based-on-spatial-fractional-order-geometry-and-spectral-spatial-low-rank-priors} replace the gradient on the HRMS and PAN by fractional-order gradients on them and proposes FOTV regularization. \cite{A-variational-pan-sharpening-with-local-gradient-constraints} considers the linear relationship between HRMS and PAN gradients on local patches to more carefully describe their relationship. \cite{High-quality-Bayesian-pansharpening} proposes a representation between HRMS and PAN by linear combinations of multi-order gradients. \cite{A-new-variational-approach-based-on-proximal-deep-injection-and-gradient-intensity-similarity-for-spatio-spectral-image-fusion} improves the spectral term proposed in \cite{SIRF-Simultaneous-satellite-image-registration-and-fusion-in-a-unified-framework} by adjusting the mean of the extended PAN. Besides, \cite{A-new-variational-approach-based-on-proximal-deep-injection-and-gradient-intensity-similarity-for-spatio-spectral-image-fusion} also proposes to use the output of a pre-trained neural network to guided the restoration of HRMS. \cite{VO+Net} proposes to apply the MRA-based model to formulate the relationship between HRMS and PAN, which achieves remarkable performance. Soon, \cite{LRTCFPan} also applies the MRA-based approach to a low-rank tensor reconstruction model. \cite{SFNLR} formulates a very concise representation of HRMS by the Hadmard product of a coefficient and an extended PAN. In \cite{VO+Net, LRTCFPan, SFNLR}, the corresponding coefficients are estimated at the beginning and fixed when solving the optimization problem. 

Applying deep neural networks to solve multispectral pansharpening has become a popular trend since the remarkable development of deep learning. Supervised DL-based pansharpening methods usually achieve superior performance due to the carefully designed network structure and the large amount of training data that contains ground-truth HRMS \cite{Pansharpening-by-convolutional-neural-networks, PanNet-A-deep-network-architecture-for-pan-sharpening, DRPNN, Pan-sharpening-using-an-efficient-bidirectional-pyramid-network,PSGAN, MSDCNN, Pansharpening-via-detail-injection-based-convolutional-neural-networks, FusionNet, P2Sharpen, Laplacian-pyramid-networks-A-new-approach-for-multispectral-pansharpening, Going-deeper-with-densely-connected-convolutional-neural-networks-for-multispectral-pansharpening, Deep-multiscale-detail-networks-for-multiband-spectral-image-sharpening, Pan-sharpening-via-multiscale-dynamic-convolutional-neural-network}. However, supervised DL-based methods may suffer performance reduction when tested on a different dataset. Besides, training on the reduced resolution data, which is most available, does not necessarily lead to the same good performance on full resolution data \cite{Machine-learning-in-pansharpening-A-benchmark-from-shallow-to-deep-networks}. Thus, unsupervised DL-based methods have been investigated \cite{Deep-image-interpolation-A-unified-unsupervised-framework-for-pansharpening, PANGAN, LDPNet, PercepPan, PGMAN,Unsupervised-MunGAN, UPanGAN, Unsupervised-pansharpening-based-on-self-attention-mechanism, Unsupervised-pansharpening-method-using-residual-network-with-spatial-texture-attention, Unsupervised-cycle-consistent-generative-adversarial-networks-for-pan-sharpening}. They are largely based on the network structure, the combination of inputs, and the design of loss functions. The training label requirement is released, but a large number of LRMS/PAN training pairs are still needed. Testing is usually performed on the same dataset. Most recently, zero-shot pansharpening methods have been proposed, in which no additional training data is required. The method in \cite{ZSPan} contains three components, that is, RSP to pre-train a network through the supervised learning way, SDE to learn the spatial relationship between HRMS and PAN, and finally FUG to train the fusion network using the results of RSP and SDE. \cite{Zero-Sharpen} propose to solve a variational optimization problem. Its objective contains the output of a neural network, which is an approximation to the HRMS. The networks are optimized by a series of loss functions independent of the variational optimization problem. Unlike \cite{Zero-Sharpen}, our method is built within one variational optimization problem, which is more concise and easier to implement. 

\subsection{DIP-based image restoration}
\cite{DIP} has made a fascinating discovery about image generation and restoration when using deep convolutional networks. That is the image priors can be sufficiently captured by the network structure itself. The network takes a random tensor as input. Its parameters are randomly initialized and then optimized by minimizing the distance between the network output and the only degraded observation. When optimization is finished, the network output is taken as the restored image. DIP has its connection with VO-based methods if seeing the network structure as the implicit regularization. Note that DIP does not require ground-truth images for training. Thus, it can be classified into the zero-shot method and usually has good generalizability. Following this idea, many image restoration methods have been proposed, e.g., medical image reconstruction \cite{Computed-tomography-reconstruction-using-deep-image-prior-and-learned-reconstruction-methods, PET-image-reconstruction-using-deep-image-prior, Time-dependent-deep-image-prior-for-dynamic-MRI}, HSI unmixing \cite{UnDIP-Hyperspectral-unmixing-using-deep-image-prior}, dehazing \cite{Learning-deep-priors-for-image-dehazing, Proximal-dehaze-net-A-prior-learning-based-deep-network-for-single-image-dehazing}, HSI denoising \cite{Hyperspectral-Denoising-Using-Unsupervised-Disentangled-Spatiospectral-Deep-Priors, Hyperspectral-mixed-noise-removal-via-spatial-spectral-constrained-unsupervised-deep-image-prior}, etc. In this work, the implicit prior induced by neural work is not applied directly to the expected image as previous works did. Instead, we use it to model the unknown relationship between HRMS and PAN.

\section{Main Method}\label{sec-Main-method}
\begin{figure*}[t]
	\centering
	\includegraphics[width=18cm]{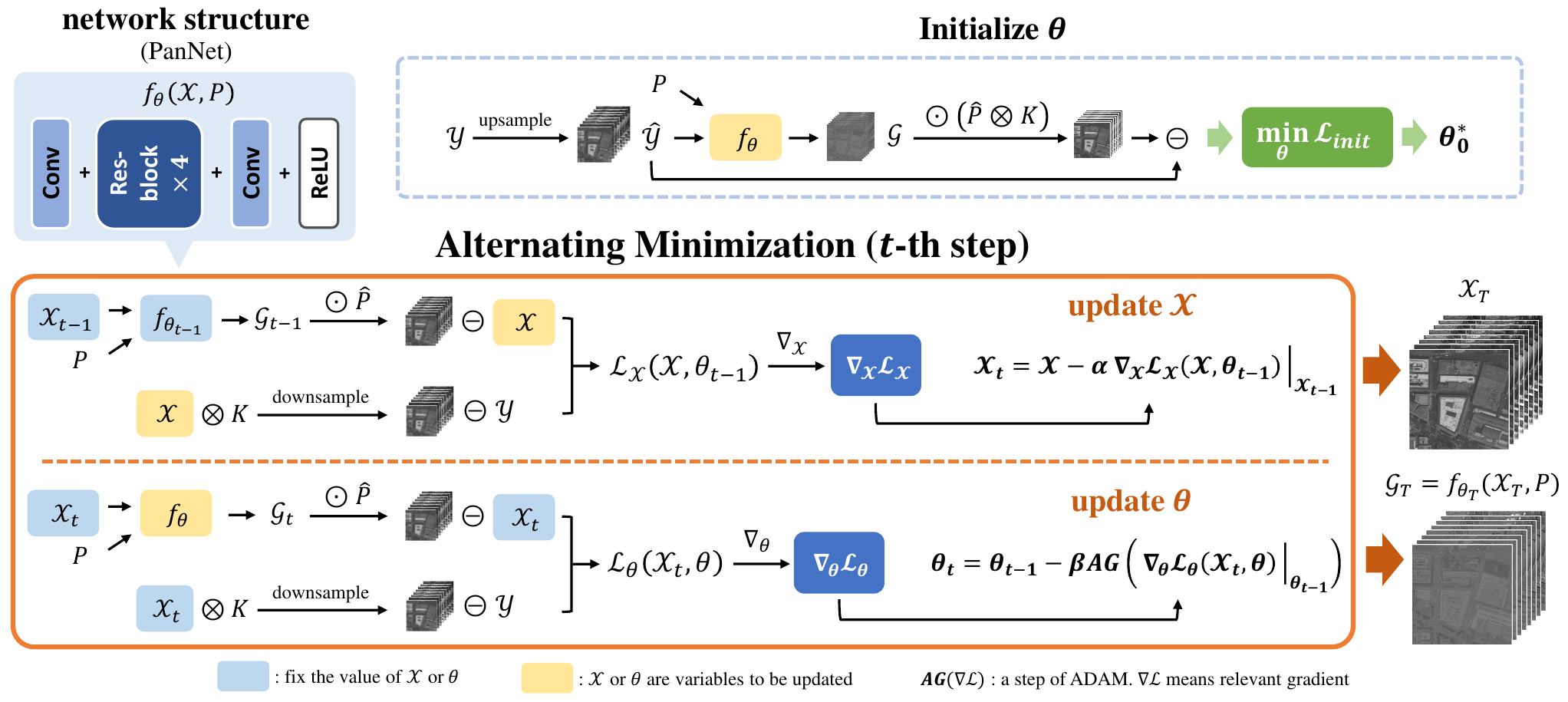}
	\caption{An overview of the proposed PSDip. We formulate an optimization problem $\min_{\X,\theta}~\|\Y - (\X\otimes K)\downarrow_r \|_F^2 + \lambda \|\X - f_\theta(\X, P)\odot\hat{P}  \|_F^2$ for multispectral pansharpening. The optimization objective is denoted as $\la(\X,\theta)$. The network $f_\theta(\X,P)$ takes the HRMS $\X$ and PAN $P$ as inputs and outputs the coefficient tensor $\G$ which comes from the representation $\X=\G\odot\hat{P}$. We choose PanNet \cite{PanNet-A-deep-network-architecture-for-pan-sharpening} as the backbone of $f_\theta$ (shown in the upper-left part). Before solving the optimization problem, we initialize $f_\theta$ by $\min_\theta ~\|\hat{\Y} - f_\theta(\hat{\Y}, P)\odot(\hat{P}\otimes K)\|$ so that $f_\theta$ predicts a rough $\G$ (shown in the upper-right part). Then, the network parameters and the expected HRMS are formally optimized in the proposed model by alternating minimization (shown in the lower part). Each subproblem is solved by gradient descent based methods. When the optimization finishes after $T$ steps, we derive the expected HRMS $\X_T$ as well as the coefficient tensor $\G_T = f_{\theta_T}(\X_T,P)$.}
	\label{fig-zhutu}
\end{figure*}
We use capital letters to represent a matrix, e.g., $A\in\R^{H\times W}$. A tensor that has more than two dimensions is written in calligraphy, e.g., $\X\in\R^{H\times W\times S}$. The $(i,j,k)$-th element of $\X$ is denoted as $\X_{ijk}$. ``$A\otimes B$" means the convolution between $A$ and $B$. ``$\odot$" is the Hadmard product, i.e., element-wise multiplication. Correspondingly, ``$\oslash$" is element-wise division. ``$A\Dn_r$" means downsampling $A$ by the scale factor of $r$. The Frobenius norm of a matrix/tenor is written as ``$\|\cdot\|_F$", i.e., $\|\X\|_F:=\sqrt{\sum_{ijk}\X_{ijk}^2}$.

\subsection{Overview to pansharpening}
Multispectral pansharpening aims to fuse the observed LRMS and PAN into the expected HRMS. Let $\X\in\R^{H\times W\times S}$ denote the HRMS, where $H,W$ and $S$ represent the height, width and spectral dimension of HRMS, respectively. The corresponding LRMS refers to as $\Y\in\R^{h\times w\times S}$ and PAN is written as $P\in\R^{H\times W}$. The downscaling factor of the LRMS is written as $r$, i.e., $H/h = W/w = r$. Taking multispectral pansharpening methods as an inverse problem \cite{Introduction-to-inverse-problems-in-imaging}, we can formulate the following model \cite{Variational-methods-in-imaging} for general VO-based methods \cite{Review}:
\begin{align}\label{pan-model}
	\min_\X~L_y(\X,\Y) + \lambda_1L_p(\X,P) + \lambda_2R(\X).
\end{align}
$L_y(\cdot, \cdot)$ and $L_p(\cdot, \cdot)$ represent the data fidelity terms to describe the relationships of HRMS/LRMS and HRMS/PAN, respectively.  $R(\X)$ means the regularization term to characterize the structure of $\X$. $\lambda_1$ and $\lambda_2$ are two trade-off parameters to balance the proportion of each term. 

The most adopted relationship between HRMS and LRMS in VO-based pansharpening methods is as follows \cite{Pansharpening-based-on-semiblind-deconvolution}:
\begin{align}
	\Y = (\X\otimes K)\Dn_r + n_1.
\end{align}
$K\in\R^{k\times k}$ denotes the blur kernel. $n_1$ means the small residual value which is usually modeled by zero-mean Gaussian distribution. In brief, the above representation means that the LRMS is considered as been blurred and downsampled by HRMS. Thus, $L_y$ can be formulated by
\begin{align}
	L_y = \left\| \Y - (\X\otimes K)\downarrow_r \right\|_F^2.
\end{align}
Following previous works, we also adopt the Gaussian filter matched with the Modulation Transfer Function (MTF) of the multispectral sensors as the kernel $K$ \cite{MTF}. The kernel is preset and fixed. It can be seen that the term $L_y$ mainly measures the spatial information loss since blurring and downsampling are performed in the spatial dimension.

Unlike $L_y$, the term $L_p$ is more difficult to design because it remains an open problem how the PAN degenerates from the HRMS. In the earliest works \cite{A-variational-model-for-P+XS-image-fusion}, the simple linear model $P = \X\times_3 \bm{\mathrm{p}}$ was proposed and then used for many other works \cite{A-new-pan-sharpening-method-using-a-compressed-sensing-technique,A-new-pansharpening-algorithm-based-on-total-variation,A-variational-pansharpening-approach-based-on-reproducible-kernel-Hilbert-space-and-heaviside-function,Single-image-super-resolution-via-an-iterative-reproducing-kernel-Hilbert-space-method,Multiband-remote-sensing-image-pansharpening-based-on-dual-injection-model}. It means performing a linear combination of the spectral bands of $\X$ with the coefficients vector $\bm{\mathrm{p}}\in\R^{S\times 1}$. One of the biggest advantages of linear HRMS/PAN model is the computational convenience when solving the optimization problem because it only contains matrix multiplication operation. Besides, the spatial information can be well included since the operation is performed on the spectral dimension. However, the linear model has several drawbacks in improving the model performance. First, even if $P = \X\times_3 \bm{\mathrm{p}}$ strictly holds, the solution space of this equation is quite large since $\mathrm{rank}(\bm{\mathrm{p}})=1$, meaning that there still exists much more image information that the linear model can not provide. Second, if the relationship $P = \X\times_3 \bm{\mathrm{p}}$ does not strictly hold, such linear model would inevitably bring approximation errors.

In order to further improve the model performance., more complex models to describe HRMS/PAN have been investigated to improve the guidance of PAN \cite{A-variational-approach-for-pan-sharpening, A-new-pansharpening-method-based-on-spatial-and-spectral-sparsity-priors,Pan-sharpening-with-a-hyper-Laplacian-penalty,SIRF-Simultaneous-satellite-image-registration-and-fusion-in-a-unified-framework,A-variational-pan-sharpening-method-based-on-spatial-fractional-order-geometry-and-spectral-spatial-low-rank-priors,A-variational-pan-sharpening-with-local-gradient-constraints, High-quality-Bayesian-pansharpening,A-new-variational-approach-based-on-proximal-deep-injection-and-gradient-intensity-similarity-for-spatio-spectral-image-fusion,VO+Net,LRTCFPan,SFNLR}. For example, a few works seek back to representations proposed by CS- and MRA-based methods \cite{VO+Net, LRTCFPan, SFNLR}. Among them, the most simple model represents HRMS by
\begin{align}\label{hrms-pan}
	\X = \G \odot \hat{P}.
\end{align}
Eq. \eqref{hrms-pan} has a very concise and flexible form, which is similar to Brovey transform \cite{Brovey}. Only two components are included in this representation. The extended PAN $\hat{P}\in\R^{H\times W\times S}$ is constructed from the PAN and has the same size, especially the spatial resolution, as the HRMS. For example, $\hat{P}$ could be histogram-matched by the PAN and LRMS. We see that in this representation, $\hat{P}$ mainly provides spatial information for HRMS. However, since PAN only has one channel, $\hat{P}$ does not preserve consistent spectral information with $\X$. That is why it needs a coefficient tensor $\G$ to balance the approximation. Theoretically, Eq. \eqref{hrms-pan} could always hold no matter what real relationship between $\X$ and $P$ is. Following Eq. \eqref{hrms-pan}, the term $L_p$ can be formulated as
\begin{align}\label{spe}
	L_p = \left\| \X - \G \odot \hat{P} \right\|_F^2.
\end{align}
Let us take a closer look to Eq. \eqref{spe}. Suppose $\G$ is completely undetermined, $L_p$ obviously can not be used for problem (\ref{pan-model}) because whatever $\X$ is, we can always set $\G=\X\oslash\hat{P}$ so that $L_p$ reaches its minimum value, i.e., zero. In other words, $\X = \G \odot \hat{P}$, if without additional constraints, actually does not reveal effective image information about $\X$. Previously, in both CS-based \cite{Brovey,GS,GSA,BDSD,PRACS}, MRA-based \cite{HPF,Indusion,GLP,SFIM} and relevant VO-based methods \cite{VO+Net, LRTCFPan, SFNLR}, $\G$ should be preset and various ingenious approaches have been proposed. When $\G$ is determined by a certain method, we see that $\G\odot\hat{P}$ is a direct approximation to $\X$ and $L_p$ directly constrains $\X$. 

\subsection{Proposed model}\label{sec-proposed-method}
Unlike existing works that mainly focus on how to preset $\G$, we consider $\G$ in Eq. \eqref{hrms-pan} as a latent variable and optimize it together with $\X$ in multispectral pansharpening problem (\ref{pan-model}). Specifically, we reformulate (\ref{pan-model}) as 
\begin{align}\label{our-model}
	\min_{\X,\G}~\left\|\Y - (\X\otimes K)\downarrow_r \right\|_F^2 + \lambda_1\left\|\X - \G\odot\hat{P}\right\|_F^2 + \lambda_2 R(\G).
\end{align}
In the proposed model \eqref{our-model}, a regularization term about $\G$ is added. As mentioned above, if the regularization is directly about $\X$, $L_p = \|\X - \G\odot\hat{P} \|_F^2$ can always reach zero no matter what $\X$ is. Then, $\hat{P}$, which contains information from PAN, does not effectively contribute to the formulated problem. However, the regularization on $\G$ makes the spatial fidelity term $L_2$ affect the problem solution. In this way, problem \eqref{our-model} is not highly ill-posed and the expected HRMS then tends to be more properly guided by both LRMS and PAN.

Besides, we simply remove $R(\X)$ in model \eqref{our-model}. Note that since $\G\odot\hat{P}$ is an approximation to $\X$, regularization on $\G$ can be seen as indirectly constraining the structure of $\X$. To see this more clearly, we may write $R(\G)\approx R(\X\oslash\hat{P})$. Furthermore, we have the following theorem that implies that $R(\X)$ can actually be ``absorbed" into $R(\G)$. 
\begin{theorem}\label{theorem}
	For any regularization $R_g(\G)$ and $R_x(\X)$, if $(\X^*, \G_1)$ is the minimum point of the following problem
	\begin{align}\label{rx-model}
		\min_{\X,\G}~ & \left\|\Y - (\X\otimes K)\downarrow_r \right\|_F^2 + \lambda_1\left\|\X - \G\odot\hat{P}\right\|_F^2 + \nonumber\\ 
		& \lambda_x R_x(\X) + \lambda_g R_g(\G),
	\end{align}
	then, there exists at least one $R(\G)$ and $\G_2$, such that $(\X^*, \G_2)$ is the minimum point of problem \eqref{our-model}.
\end{theorem}
The proof is presented in Appendix \ref{appendix}. With Theorem \ref{theorem}, our model \eqref{our-model} is somehow equivalent to model \eqref{rx-model} with respect to $\X$. Comparing these two models, its seems like that $R_x(\X)$ can be ``absorbed" into $R_g(\G)$ to form a new $R(\G)$.
Since the main focus of this work is about $\G$, we remove the regularization $R(\X)$ to keep the model \eqref{our-model} as compact and concise as possible. Also we want to make our algorithm possibly easily reproducible. Considering this point, we prefer to keep the form of model \eqref{our-model} in this study.

\begin{figure}[t]
	\newcommand{\mysize}{2.8cm}
	\centering
	\begin{minipage}[t]{\mysize}
		\centering
		\small
		\includegraphics[width=\mysize]{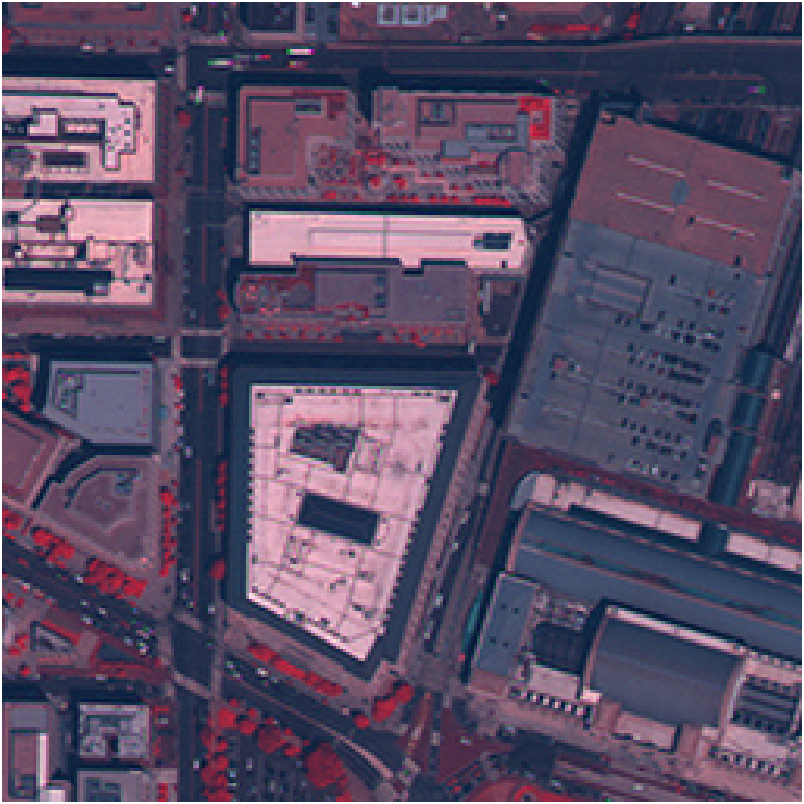} \\
		HRMS $\X$
	\end{minipage}
	\begin{minipage}[t]{\mysize}
		\centering
		\small
		\includegraphics[width=\mysize]{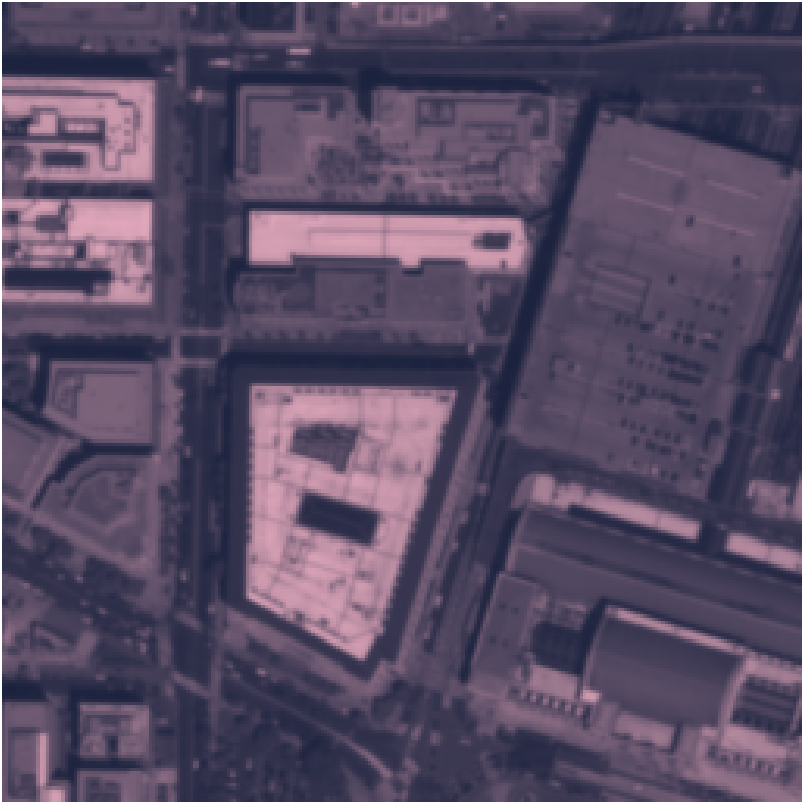} \\
		$\hat{P}$
	\end{minipage}
	\begin{minipage}[t]{\mysize}
		\centering
		\small
		\includegraphics[width=\mysize]{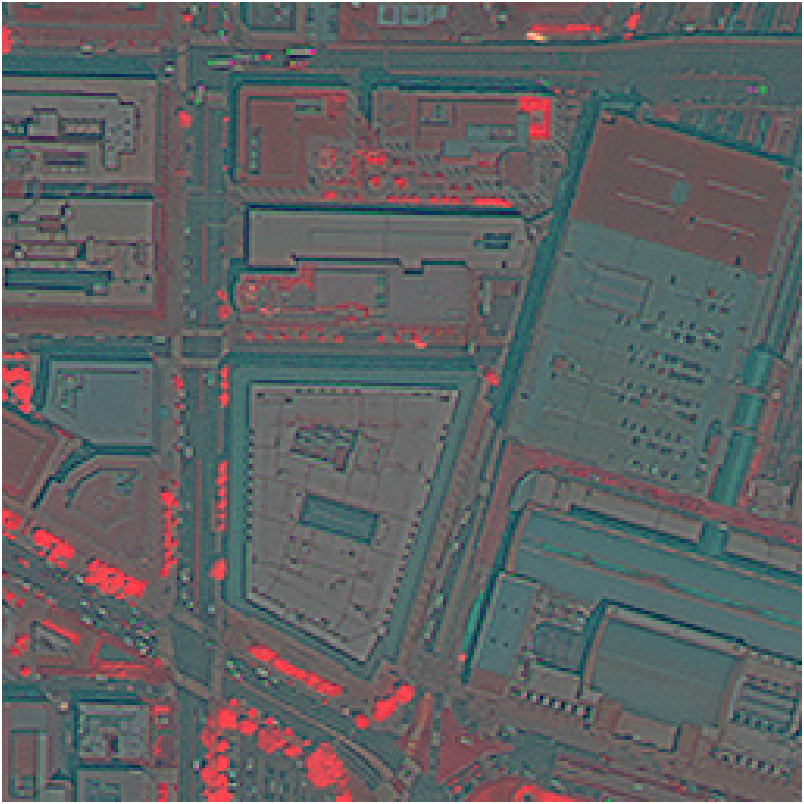} \\
		$\G = \X\oslash\hat{P}$
	\end{minipage}
	\caption{Pseudo-color images of HRMS $\X$, the extended PAN $\hat{P}$ and the coefficient $\G$.}
	\label{fig-anaG}
\end{figure}

As revealed by \cite{SFNLR}, the coefficient tensor $\G$ contains image structures although it is not strictly an ``image". An example is shown in Fig. \ref{fig-anaG}. We can see clear textures existed in $\G$. These textures contain not only spatial details which make $\G$ look like an image but also spectrum compensation information to match $\hat{P}$ with $\X$. The presence of these structures in $\G$ is mainly attributable to the high spatial similarity between HRMS and PAN. Specifically, since the spatial resolution of HRMS and $\hat{P}$ should be the same, the positions of the low-frequency area and the high-frequency area between HRMS and $\hat{P}$ are nearly the same. Thus, the element-division result of them, i.e., $\G$, would give rise to similar ``low-frequency" and ``high-frequency" structures in the same positions. Based on this observation, a direct choice for $R(\G)$ is regularization used for images. In this work, we propose to use a neural network $f_\theta(\cdot)$ to estimate the coefficient tensor $\G$. Following the idea of DIP \cite{DIP}, the network structure itself can implicitly regularize its output with implicit image prior. Then, we derive the proposed method which has the following optimization objective
\begin{align}\label{main-method}
	\min_{\X,\theta}~\left\|\Y - (\X\otimes K)\downarrow_r \right\|_F^2 + \lambda \left\|\X - f_\theta(\X, P)\odot\hat{P}  \right\|_F^2,
\end{align}
where we have $\G = f_\theta(\X,P)$. The regularization $R(\G)$ is absorbed in the network $f_\theta(\cdot)$ and does not appear explicitly. For the general setting of DIP \cite{DIP}, $f_\theta$ would take random noise as input. In this way, the network learns to construct the target purely from ``nothing". Here in problem (\ref{main-method}), the network $f_\theta$ is designed to take $\X$ and $P$ as inputs. This is for three reasons. First, the representation $\X = \G\odot\hat{P}$ shows an evident relationship between $\X,\G$ and $P$. Thus, it is reasonable to consider constructing a mapping from $\X$ and $P$ to $\G$. Second, these two inputs could provide additional information for $f_\theta$ to construct $\G$ except for the network structure, which we empirically find to be very useful to get better estimation. Third, we can dynamically and gradually modify $\G$ by adjusting the input $\X$ except for only optimizing the network parameters. We call the proposed problem (\ref{main-method}) \textbf{PSDip}.

\subsection{Optimization for the PSDip model}\label{sec-solve-the-proposed-method}
PSDip can be conveniently solved by alternating minimization \cite{Numerical-optimization}\cite{Machine-learning-a-probabilistic-perspective}. Let $\la(\X,\theta)$ denote the objective function of problem (\ref{main-method}) 
\begin{align}
	\la(\X, \theta)\! := \!\left\|\Y \!-\! (\X\otimes K)\downarrow_r \right\|_F^2\! +\! \lambda \left\|\X\! -\! f_\theta(\X, P)\odot\hat{P}  \right\|_F^2.
\end{align}
In the $t$-th step, $\X$ should be updated by solving the corresponding $\X$-subproblem where the network parameter $\theta$ is fixed in $\la(\X,\theta)$. The subproblem does not have a closed-form solution. Thus, we consider updating $\X$ by applying one step of gradient descent. Besides, we empirically find that the algorithm does not produce the best results if the network input ``$\X$" is also treated as the variable in the $\X$-subproblem, i.e., the gradient with respect to network input $\X$ is also contributed to update $\X$ in this step. Details are presented in Sec. \ref{sec-ana-X}. Thus, we further simplify the $\X$-subproblem in the $t$-th step by fixing the network input $\X$ as the last updated one, i.e., $\X_{t-1}$. In this way, the simplified $\X$-subproblem in the $t$-th step has the following objective:
\begin{align}
	&\la_\X(\X, \theta_{t-1}) := \nonumber \\
	& \left\|\Y \!-\! (\X\otimes K)\downarrow_r \right\|_F^2\! +\! \lambda \left\|\X\! -\! f_{\theta_{t-1}}(\X_{t-1}, P)\odot\hat{P}  \right\|_F^2.
\end{align}
Then, $\X$ can be simply updated by gradient descent:
\begin{align}\label{update-X}
	\X_{t} = \X_{t-1} - \left.\alpha\nabla_\X \la_\X(\X,\theta_{t-1})\right|_{\X_{t-1}},
\end{align}
where $\alpha$ is the step size. Network parameter $\theta$ is updated by solving the corresponding $\theta$-subproblem where $\X$ is fixed in $\la(\X,\theta)$. Specifically, the objective of $\theta$-subproblem in the $t$-th step has the form of
\begin{align}
	& \la_\theta(\X_t,\theta):= \nonumber \\
	& \left\|\Y - (\X_t\otimes K)\downarrow_r \right\|_F^2\! + \lambda \left\|\X_t\! - f_\theta(\X_t, P)\odot\hat{P}  \right\|_F^2.
\end{align}
We use Adam \cite{Adam} to update $\theta$ as most DL-based methods do. Similar to $\X_t$, $\theta_t$ is calculated by one step updating:
\begin{align}\label{update-theta}
	\theta_t = \theta_{t-1} - \beta\mathrm{AG}\left(\nabla_\theta \la(\X_t,\theta)|_{\theta_{t-1}}\right),
\end{align}
where $\mathrm{AG}(\cdot)$ means the update direction in Adam\footnote{For the sake of convenience, only the gradient w.r.t. $\theta$ is shown in $\mathrm{AG}$ but one should know that $\mathrm{AG}$ includes complete update components (e.g. moments) of one step in Adam.} and $\beta$ is the learning rate. 

A good initial value of $\theta$ for the alternating minimization could help to stabilize the updating process and then help our model \eqref{our-model} to achieve better performance. We see that applying the blurring operator on both sides of $\X \approx \G\odot \hat{P}$ gives rise to $\X\otimes K \approx \G\odot (\hat{P}\otimes K)$ \cite{SFNLR}. Thus, we initialize $\theta$ by 
\begin{align}\label{init-theta}
	\theta_0^* = \arg\min_\theta ~\left\|\hat{\Y} - f_\theta(\hat{\Y}, P)\odot(\hat{P}\otimes K)\right\|_F,
\end{align}
where $\hat{\Y}\approx \X\otimes K$ means the upsampled LRMS. $\hat{\Y}$ is also considered as the first input of $f_\theta$ since we do not have access to $\X$. Then, $\theta_0^*$ is used as the initial value of $\theta$ for alternating minimization for the problem (\ref{main-method}). In Algorithm \ref{algo-my}, we summarize the entire process of implementing our PSDip. Implementation details about Algorithm \ref{algo-my} are presented in Sec. \ref{sec-imple-details}. In addition, Fig. \ref{fig-zhutu} shows the PSDip flow chart to take a comprehensive look at PSDip.

\begin{algorithm}[t]
	\renewcommand{\algorithmicrequire}{\textbf{Input:}}
	\renewcommand{\algorithmicensure}{\textbf{Initialization:}}
	\renewcommand{\algorithmicreturn}{\textbf{Output:}}
	\caption{Proposed Method PSDip}
	\label{algo-my}
	\begin{algorithmic}[1]
		\REQUIRE{LRMS $\Y$. Extended PAN $\hat{P}$. Network $f_\theta(\cdot)$. Trade-off parameters $\lambda$. Learning rate $\alpha$ and $\beta$. Iteration steps $T$.}
		\ENSURE{$\X_0=\hat{\Y}$, $\theta_0^*$ by Eq. \eqref{init-theta}}\;	
		\FOR{$t = 1:T$}
		\STATE update $\X_t$ by Eq. \eqref{update-X} 
		\STATE update $\theta_t$ by Eq. \eqref{update-theta}
		\ENDFOR
		\RETURN{$\X_{T}, \G_T=f_{\theta_T}(\X_T,P)$}
	\end{algorithmic}
\end{algorithm}

\subsection{Implementation details}\label{sec-imple-details}
We adopt PanNet \cite{PanNet-A-deep-network-architecture-for-pan-sharpening} as the backbone of our network $f_\theta$. The network mainly contains convolutional layers and skip connections. In addition, we add ReLU activation to the final layer of PanNet to ensure that the output is always positive. It should be noted that the network structure is not exclusively specified. In Sec. \ref{sec-disG}, we also present the results of two other networks. To construct $\hat{P}$, we first perform histogram-matching to generate $\hat{P}'$, i.e., each band of $\hat{P}'$ is entirely translated and stretched such that the mean and standard values match those of the same band of LRMS. To avoid zero-value in denominator, we then add a small value ($\varepsilon = 1e-2$) to $\hat{P}'$ and finally get $\hat{P} = \hat{P}' + \varepsilon$. The upsampled LRMS $\hat{\Y}$ is derived by general bicubic interpolation of LRMS $\Y$. The kernel $K$ matches the MTF of multispectral sensors \cite{MTF}. For the initialization problem (\ref{init-theta}), we use Adam to optimize $\theta$ until the objective converges, which takes about 8000 steps. The learning rate is constantly set as $\mathrm{1e}^{-3}$. For the alternating minimization for the main problem (\ref{main-method}), we set $\alpha=2$ and $\beta=\mathrm{1e^{-3}}$ for all experiments. In each step of alternating minimization, $\X$ and $\theta$ are updated by one step of gradient descent and ADAM, respectively. The trade-off parameter $\lambda$ is set as 0.1 for all experiments. The iteration is also completed when the value of $\la(\X,\theta)$ changes slowly, which takes about 3000 steps. That is, we set $T=3000$ in Algorithm \ref{algo-my}.

\section{Experiments}\label{sec-experiments}

\begin{table*}[t]
	\renewcommand{\arraystretch}{1.29}
	\newcommand{\mysize}{1.9cm}
	\fontsize{9}{10}\selectfont
	\caption{Test performance on the reduced resolution WV2 dataset. ``T" means running time. The best results are in \textbf{bold}, and the second best results are with \ul{underline}.}
	\label{tab-WV2}
	\centering
	\begin{tabular}{ L{2.1cm} | M{\mysize+0.2cm} M{\mysize} M{\mysize} M{\mysize+0.2cm} M{\mysize+0.2cm} M{\mysize} | M{0.6cm}}
		\Xhline{1pt}
		Methods & PSNR$\uparrow\pm$ std & SSIM$\uparrow\pm$ std & Q8$\uparrow\pm$ std &         SAM$\downarrow\pm$ std & ERGAS$\downarrow\pm$ std & SCC$\uparrow\pm$ std & \text{T (s)}  \\ 
		\hline 
		BT-H \cite{BT-H}& 30.160 $\pm$ 1.919 & 0.832 $\pm$ 0.032 & 0.331 $\pm$ 0.104 & 5.891 $\pm$ \ul{0.765} & 4.398 $\pm$ \ul{0.590} & \ul{0.917} $\pm$ \ul{0.010} & 0.09 \\ 
		BDSD-PC \cite{BDSD-PC}& 30.587 $\pm$ \tb{1.783} & 0.840 $\pm$ 0.030 & 0.355 $\pm$ 0.103 & 6.143 $\pm$ 0.914 & 4.253 $\pm$ 0.717 & 0.912 $\pm$ 0.018 & 0.73 \\ 
		C-GSA \cite{C-GSA}& 30.129 $\pm$ 1.880 & 0.814 $\pm$ 0.036 & 0.306 $\pm$ 0.106 & 6.249 $\pm$ 0.880 & 4.521 $\pm$ 0.665 & 0.892 $\pm$ 0.021 & 1.15 \\ 
		AWLP \cite{AWLP}& 30.081 $\pm$ 1.917 & 0.830 $\pm$ 0.033 & 0.369 $\pm$ \tb{0.091} & 6.068 $\pm$ 0.791 & 4.437 $\pm$ 0.609 & 0.911 $\pm$ 0.010 & 0.10 \\ 
		GLP-HPM \cite{GLP-HPM} & 30.176 $\pm$ 1.918 & 0.816 $\pm$ 0.040 & 0.315 $\pm$ 0.109 & 6.352 $\pm$ 0.960 & 4.555 $\pm$ 0.781 & 0.890 $\pm$ 0.037 & 0.12 \\ 
		GLP-FS \cite{GLP-FS}& 30.187 $\pm$ \ul{1.833} & 0.818 $\pm$ 0.037 & 0.306 $\pm$ 0.110 & 6.187 $\pm$ 0.935 & 4.454 $\pm$ 0.704 & 0.894 $\pm$ 0.024 & 0.12 \\ 
		LDP \cite{LDPNet}& 28.659 $\pm$ 2.002 & 0.762 $\pm$ 0.050 & 0.221 $\pm$ 0.124 & 8.811 $\pm$ 6.032 & 5.814 $\pm$ 2.040 & 0.861 $\pm$ 0.038 & 871.1 \\ 
		LRTCFPan \cite{LRTCFPan}& \ul{30.765} $\pm$ 1.936 & \ul{0.843} $\pm$ \ul{0.029} & \ul{0.370} $\pm$ 0.100 & \ul{5.622} $\pm$ 0.767 & \ul{4.117} $\pm$ 0.633 & 0.914 $\pm$ 0.012 & 55.47 \\ 
		ZSPan \cite{ZSPan}& 29.841 $\pm$ 1.981 & 0.826 $\pm$ 0.033 & 0.346 $\pm$ 0.102 & 6.037 $\pm$ 0.845 & 4.554 $\pm$ 0.666 & 0.905 $\pm$ 0.011 & 43.96 \\ 
		\hline
		PSDip & \tb{31.174} $\pm$ 1.985 & \tb{0.860} $\pm$ \tb{0.028} & \tb{0.421} $\pm$ \ul{0.086} & \tb{5.529} $\pm$ \tb{0.722} & \tb{3.896} $\pm$ \tb{0.531} & \tb{0.931} $\pm$ \tb{0.008} & 279.0 \\ 
		\Xhline{1pt}
	\end{tabular}
\end{table*}

\begin{table*}[t]
	\renewcommand{\arraystretch}{1.29}
	\newcommand{\mysize}{1.9cm}
	\fontsize{9}{10}\selectfont
	\caption{Test performance on the reduced resolution WV3 dataset. ``T" means running time. The best results are in \textbf{bold}, and the second best results are with \ul{underline}.}
	\label{tab-WV3}
	\centering
	\begin{tabular}{ L{2.1cm} | M{\mysize+0.2cm} M{\mysize} M{\mysize} M{\mysize+0.2cm} M{\mysize+0.2cm} M{\mysize} | M{0.6cm}}
		\Xhline{1pt}
		Methods & PSNR$\uparrow\pm$ std & SSIM$\uparrow\pm$ std & Q8$\uparrow\pm$ std &         SAM$\downarrow\pm$ std & ERGAS$\downarrow\pm$ std & SCC$\uparrow\pm$ std & \text{T (s)}  \\ 
		\hline 
		BT-H \cite{BT-H}& 33.098 $\pm$ 2.744 & 0.896 $\pm$ 0.026 & 0.494 $\pm$ 0.166 & 4.873 $\pm$ 1.379 & 4.550 $\pm$ 1.456 & 0.925 $\pm$ 0.024 & 0.02 \\ 
		BDSD-PC \cite{BDSD-PC}& 32.972 $\pm$ 2.654 & 0.890 $\pm$ 0.031 & 0.500 $\pm$ 0.170 & 5.402 $\pm$ 1.775 & 4.677 $\pm$ 1.579 & 0.907 $\pm$ 0.040 & 0.07 \\ 
		C-GSA \cite{C-GSA}& 32.769 $\pm$ 2.805 & 0.876 $\pm$ 0.030 & 0.467 $\pm$ 0.174 & 5.543 $\pm$ 1.651 & 4.815 $\pm$ 1.525 & 0.898 $\pm$ 0.037 & 0.27 \\ 
		AWLP \cite{AWLP}& 32.874 $\pm$ 2.827 & 0.891 $\pm$ 0.025 & \ul{0.566} $\pm$ \tb{0.114} & 5.185 $\pm$ 1.494 & 4.632 $\pm$ 1.456 & 0.913 $\pm$ 0.030 & 0.09 \\ 
		GLP-HPM \cite{GLP-HPM}& 32.999 $\pm$ 2.672 & 0.887 $\pm$ 0.032 & 0.485 $\pm$ 0.175 & 5.297 $\pm$ 1.715 & 5.130 $\pm$ 2.747 & 0.891 $\pm$ 0.113 & 0.09 \\ 
		GLP-FS \cite{GLP-FS}& 32.968 $\pm$ 2.621 & 0.884 $\pm$ 0.032 & 0.477 $\pm$ 0.181 & 5.279 $\pm$ 1.719 & 4.678 $\pm$ 1.558 & 0.901 $\pm$ 0.045 & 0.11 \\ 
		LDP \cite{LDPNet}& 30.340 $\pm$ \tb{2.422} & 0.824 $\pm$ 0.055 & 0.424 $\pm$ 0.165 & 11.014 $\pm$ 9.162 & 6.693 $\pm$ 2.853 & 0.846 $\pm$ 0.079 & 890.2 \\ 
		LRTCFPan \cite{LRTCFPan}& \ul{33.627} $\pm$ 2.699 & \ul{0.903} $\pm$ \ul{0.025} & 0.529 $\pm$ 0.150 & \ul{4.698} $\pm$ \ul{1.368} & \ul{4.288} $\pm$ \ul{1.398} & \ul{0.927} $\pm$ \ul{0.023} & 53.81 \\ 
		ZSPan \cite{ZSPan}& 32.638 $\pm$ 2.712 & 0.891 $\pm$ 0.025 & 0.508 $\pm$ 0.160 & 5.269 $\pm$ 1.419 & 4.779 $\pm$ 1.698 & 0.915 $\pm$ 0.028 & 44.78 \\ 
		\hline
		PSDip & \tb{34.395} $\pm$ \ul{2.520} & \tb{0.920} $\pm$ \tb{0.019} & \tb{0.600} $\pm$ \ul{0.120} & \tb{4.484} $\pm$ \tb{1.266} & \tb{3.842} $\pm$ \tb{1.180} & \tb{0.947} $\pm$ \tb{0.016} & 269.6 \\
		\Xhline{1pt}
	\end{tabular}
\end{table*}

\begin{table*}[t]
	\renewcommand{\arraystretch}{1.29}
	\newcommand{\mysize}{1.9cm}
	\fontsize{9}{10}\selectfont
	\caption{Test performance on the reduced resolution QB dataset. ``T" means running time. The best results are in \textbf{bold}, and the second best results are with \ul{underline}.}
	\label{tab-QB}
	\centering
	\begin{tabular}{ L{2.1cm} | M{\mysize+0.2cm} M{\mysize} M{\mysize} M{\mysize+0.2cm} M{\mysize+0.2cm} M{\mysize} | M{0.6cm}}
		\Xhline{1pt}
		Methods & PSNR$\uparrow\pm$ std & SSIM$\uparrow\pm$ std & Q4$\uparrow\pm$ std &         SAM$\downarrow\pm$ std & ERGAS$\downarrow\pm$ std & SCC$\uparrow\pm$ std & \text{T (s)}  \\ 
		\hline 
		BT-H \cite{BT-H}& 32.556 $\pm$ 3.027 & 0.867 $\pm$ 0.036 & 0.681 $\pm$ 0.138 & 7.211 $\pm$ 1.512 & 7.457 $\pm$ 0.738 & 0.915 $\pm$ 0.016 & 0.02 \\ 
		BDSD-PC \cite{BDSD-PC}& 32.462 $\pm$ 2.978 & 0.860 $\pm$ 0.036 & 0.704 $\pm$ 0.126 & 8.102 $\pm$ 1.941 & 7.567 $\pm$ 0.707 & 0.905 $\pm$ 0.017 & 0.03 \\ 
		C-GSA \cite{C-GSA}& 32.623 $\pm$ 2.959 & 0.868 $\pm$ 0.035 & 0.693 $\pm$ 0.133 & 7.259 $\pm$ 1.567 & 7.429 $\pm$ 0.702 & 0.912 $\pm$ 0.017 & 0.14 \\ 
		AWLP \cite{AWLP}& 32.402 $\pm$ 3.097 & 0.858 $\pm$ 0.039 & 0.699 $\pm$ 0.132 & 8.210 $\pm$ 1.986 & 7.629 $\pm$ 0.881 & 0.904 $\pm$ \ul{0.012} & 0.05 \\ 
		GLP-HPM \cite{GLP-HPM}& 32.520 $\pm$ \tb{1.936} & 0.868 $\pm$ \ul{0.031} & 0.698 $\pm$ 0.123 & 7.781 $\pm$ 1.745 & 9.780 $\pm$ 8.126 & 0.860 $\pm$ 0.176 & 0.04 \\ 
		GLP-FS \cite{GLP-FS}& 32.636 $\pm$ 2.836 & 0.862 $\pm$ 0.034 & 0.687 $\pm$ 0.134 & 7.801 $\pm$ 1.776 & 7.414 $\pm$ \ul{0.660} & 0.902 $\pm$ 0.024 & 0.04 \\ 
		LDP \cite{LDPNet}& 31.379 $\pm$ 3.362 & 0.818 $\pm$ 0.086 & 0.615 $\pm$ 0.169 & 10.207 $\pm$ 6.726 & 8.893 $\pm$ 2.588 & 0.885 $\pm$ 0.024 & 653.5 \\ 
		LRTCFPan \cite{LRTCFPan}& \ul{33.171} $\pm$ 3.030 & \ul{0.875} $\pm$ 0.035 & 0.710 $\pm$ 0.127 & \ul{7.207} $\pm$ 1.671 & \ul{6.977} $\pm$ 0.725 & \ul{0.916} $\pm$ 0.013 & 28.71 \\ 
		ZSPan \cite{ZSPan}& 32.376 $\pm$ \ul{2.812} & 0.865 $\pm$ 0.033 & \ul{0.722} $\pm$ \ul{0.122} & 7.772 $\pm$ \tb{1.494} & 7.578 $\pm$ 0.696 & 0.908 $\pm$ 0.016 & 33.90 \\ 
		\hline
		PSDip & \tb{34.121} $\pm$ 2.835 & \tb{0.893} $\pm$ \tb{0.028} & \tb{0.759} $\pm$ \tb{0.106} & \tb{6.744} $\pm$ \ul{1.511} & \tb{6.263} $\pm$ \tb{0.622} & \tb{0.940} $\pm$ \tb{0.012} & 209.3 \\ 
		\Xhline{1pt}
	\end{tabular}
\end{table*}

\begin{table*}[t]
	\renewcommand{\arraystretch}{1.29}
	\newcommand{\mysize}{2.2cm}
	\fontsize{9}{10}\selectfont
	\caption{Test performance on the QB full resolution dataset. ``T" means running time. The best results are in \textbf{bold}, and the second best results are with \ul{underline}.}
	\label{tab-QB-full}
	\centering
	\begin{tabular}{ L{2.3cm} | M{\mysize} M{\mysize} M{\mysize} | M{1cm}}
		\Xhline{1pt}
		Methods & QNR$\uparrow\pm$ std & $D_\lambda\downarrow\pm$ std & $D_s\downarrow\pm$ std & \text{T (s)}  \\ 
		\hline 
		BT-H \cite{BT-H} & 0.7702 $\pm$ \tb{0.0154} & 0.0499 $\pm$ 0.0135 & 0.1893 $\pm$ \tb{0.0144} & 0.08 \\ 
		BDSD-PC \cite{BDSD-PC} & 0.8059 $\pm$ 0.0510 & 0.0337 $\pm$ 0.0175 & 0.1663 $\pm$ 0.0469 & 0.07 \\ 
		C-GSA \cite{C-GSA} & 0.7581 $\pm$ 0.0297 & 0.0499 $\pm$ 0.0112 & 0.2021 $\pm$ 0.0290 & 0.98 \\ 
		AWLP \cite{AWLP} & 0.8102 $\pm$ 0.0319 & 0.0570 $\pm$ \ul{0.0103} & 0.1409 $\pm$ 0.0287 & 0.42 \\ 
		GLP-HPM \cite{GLP-HPM} & 0.8176 $\pm$ 0.0286 & 0.0499 $\pm$ 0.0114 & 0.1395 $\pm$ 0.0239 & 0.16 \\ 
		GLP-FS \cite{GLP-FS} & 0.7983 $\pm$ 0.0307 & 0.0587 $\pm$ 0.0141 & 0.1522 $\pm$ 0.0244 & 0.17 \\ 
		LDPNet \cite{LDPNet} & 0.7245 $\pm$ 0.0822 & 0.0870 $\pm$ 0.0692 & 0.2090 $\pm$ 0.0418 & 1589.3 \\ 
		LRTCFPan \cite{LRTCFPan} & \ul{0.9066} $\pm$ 0.0472 & \ul{0.0264} $\pm$ 0.0167 & \ul{0.0694} $\pm$ 0.0347 & 132.44 \\ 
		ZSPan \cite{ZSPan} & 0.8867 $\pm$ 0.0316 & 0.0341 $\pm$ 0.0193 & 0.0823 $\pm$ \ul{0.0183} & 46.16 \\ 
		\hline 
		PSDip & \tb{0.9102} $\pm$ \ul{0.0238} & \tb{0.0235} $\pm$ \tb{0.0074} & \tb{0.0680} $\pm$ 0.0184 & 349.2 \\ 
		\Xhline{1pt}
	\end{tabular}
\end{table*}

In this section, we conduct several experiments to verify the effectiveness of the proposed PSDip. The following benchmark datasets are utilized, which can be downloaded from the website\footnote{\url{https://liangjiandeng.github.io/PanCollection.html}}. Specifically, we conduct reduced resolution experiments on WorldView-2 (WV2), WorldView-3 (WV3) and QuickBird (QB) test datasets. Each of them contains 20 sets of images, in which the HRMS, LRMS and PAN are included. The HRMS sizes of WV2, WV3 and QB are $256\times 256\times 8, 256\times 256\times 8$ and $256\times 256\times 4$, respectively. The downscale factor $r$ is 4 for all datasets. That is the LRMS has the spatial size of $64\times64$. We conduct full-resolution experiments on the QB test dataset, which also contains 20 sets of images but no HRMS is available. The full-resolution PAN in the QB dataset has the size of $512\times 512$ and the corresponding LRMS has the size of $128\times128\times 4$. Besides, five images from the WV3 validation dataset are used for analyzing the trade-off parameter $\lambda$ and the setting of $f_\theta$. The original image value of each dataset ranges from 0 to 2$^{11}$. We normalize the data into [0,1] before the experiments. 

Nine typical multispectral pansharpening methods are utilized for comparison, i.e., BT-H \cite{BT-H}, BDSD-PC \cite{BDSD-PC}, C-GSA \cite{C-GSA}, AWLP\cite{AWLP}, GLP-HPM \cite{GLP-HPM}, GLP-FS \cite{GLP-FS}, LDPNet \cite{LDPNet}\footnote{\url{https://github.com/suifenglian/LDP-Net}}, LRTCFPan \cite{LRTCFPan}\footnote{\url{https://github.com/zhongchengwu/code_LRTCFPan}} and ZSPan \cite{ZSPan}\footnote{\url{https://github.com/coder-qicao/ZS-Pan}}. Specifically, BT-H, BDSD-PC and C-GSA are CS-based methods. AWLP, GLP-HPM and GLP-FS are MRA-based methods. LRTCFPan \cite{LRTCFPan} is a SOTA VO-based method. LDPNet \cite{LDPNet} is an unsupervised DL method. And ZSPan \cite{ZSPan} is a zero-shot method. For the VO-based method, the parameter setting follows the authors' suggestions in their released codes. Original LDPNet requires training the network using many LRMS/PAN pairs before testing on other LRMS/PAN pairs that belong to the same dataset. Under the zero-shot setting, we use only one pair of LRMS/PAN for both training and testing \cite{ZSPan}. Our method contains one trade-off parameter $\lambda$. For all experiments, it is set as $0.1$. All experiments are implemented in Matlab R2023a in a Computer with Inter(R) Core(TM) i7-7700 CPU and one NVIDIA GeForce RTX 3090.

We adopt the following widely-used metrics to quantitatively evaluate the restoration performance. For reduced resolution images, peak signal-to-noise ratio (PSNR), Structural Similarity Index Measure (SSIM), Q2$^n$, spectral Angle Mapper (SAM), Error Relative Global Dimension Synthesis (ERGAS) and spatial correlation coefficient (SCC) are used. For full resolution evaluation, the quality with no reference (QNR) is used \cite{Full-resolution-quality-assessment-of-pansharpening-Theoretical-and-hands-on-approaches}. This index also includes a spectral distortion index ($D_\lambda$) and a spatial distortion index ($D_s$).

\begin{figure*}[t]
	\newcommand{\mysize}{2.7cm}
	\newcommand{\name}{WV2}
	\newcommand{\suojian}{-4pt}
	\centering
	\begin{minipage}[t]{\mysize*2}
		\centering
		\small
		\includegraphics[width=\mysize]{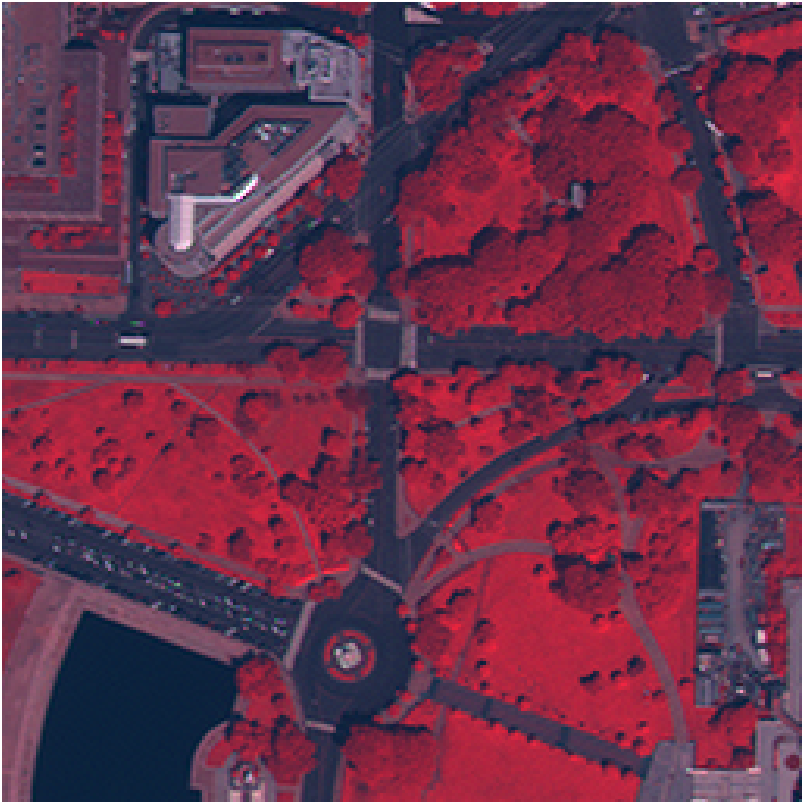}\includegraphics[width=\mysize]{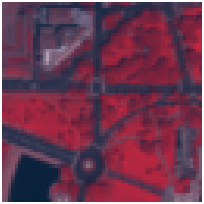} \\
		HRMS / LRMS
	\end{minipage}\hspace{\suojian}
	\foreach \i in {BT-H,BDSD-PC}
	{
		\begin{minipage}[t]{\mysize*2}
			\centering
			\small
			\includegraphics[width=\mysize]{pic//show_\name_\i.png}\includegraphics[width=\mysize]{pic//show_\name_res\i.png} \\
			\i ~/ residual \vspace{4pt}
		\end{minipage}\hspace{\suojian}
	}
	\\
	\foreach \i in {C-GSA, AWLP,GLP-HPM}
	{
		\begin{minipage}[t]{\mysize*2}
			\centering
			\small
			\includegraphics[width=\mysize]{pic//show_\name_\i.png}\includegraphics[width=\mysize]{pic//show_\name_res\i.png} \\
			\i ~/ residual \vspace{4pt}
		\end{minipage}\hspace{\suojian}
	}
	\\
	\foreach \i in {GLP-FS,LDPNet, LRTCFPan}
	{
		\begin{minipage}[t]{\mysize*2}
			\centering
			\small
			\includegraphics[width=\mysize]{pic//show_\name_\i.png}\includegraphics[width=\mysize]{pic//show_\name_res\i.png} \\
			\i ~/ residual \vspace{4pt}
		\end{minipage}\hspace{\suojian}
	}
	\\
	\foreach \i in {ZSPan,Ours}
	{
		\begin{minipage}[t]{\mysize*2}
			\centering
			\small
			\includegraphics[width=\mysize]{pic//show_\name_\i.png}\includegraphics[width=\mysize]{pic//show_\name_res\i.png} \\
			\i ~/ residual \vspace{4pt}
		\end{minipage}\hspace{\suojian}
	}
	\caption{Visualization results on WV2 reduced resolution dataset of all compared methods and the proposed PSDip. Both the restored HRMS and the residual image are shown.}
	\label{fig-WV2-reduced}
\end{figure*}

\begin{figure*}[t]
	\newcommand{\mysize}{2.7cm}
	\newcommand{\name}{WV3}
	\newcommand{\suojian}{-4pt}
	\centering
	\begin{minipage}[t]{\mysize*2}
		\centering
		\small
		\includegraphics[width=\mysize]{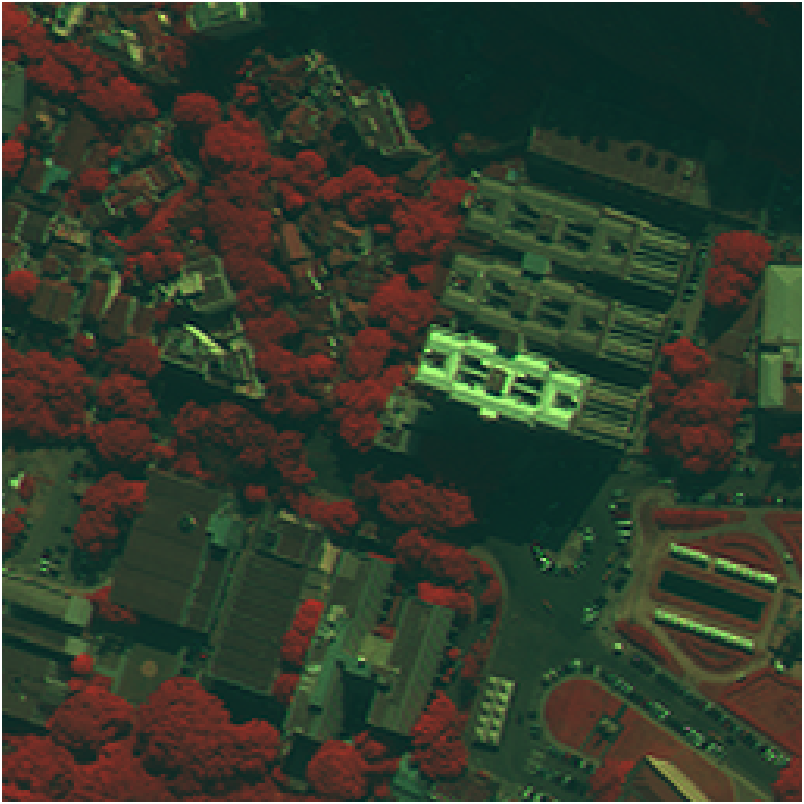}\includegraphics[width=\mysize]{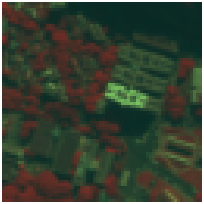} \\
		HRMS / LRMS
	\end{minipage}\hspace{\suojian}
	\foreach \i in {BT-H,BDSD-PC}
	{
		\begin{minipage}[t]{\mysize*2}
			\centering
			\small
			\includegraphics[width=\mysize]{pic//show_\name_\i.png}\includegraphics[width=\mysize]{pic//show_\name_res\i.png} \\
			\i ~/ residual \vspace{4pt}
		\end{minipage}\hspace{\suojian}
	}
	\\
	\foreach \i in {C-GSA, AWLP,GLP-HPM}
	{
		\begin{minipage}[t]{\mysize*2}
			\centering
			\small
			\includegraphics[width=\mysize]{pic//show_\name_\i.png}\includegraphics[width=\mysize]{pic//show_\name_res\i.png} \\
			\i ~/ residual \vspace{4pt}
		\end{minipage}\hspace{\suojian}
	}
	\\
	\foreach \i in {GLP-FS,LDPNet, LRTCFPan}
	{
		\begin{minipage}[t]{\mysize*2}
			\centering
			\small
			\includegraphics[width=\mysize]{pic//show_\name_\i.png}\includegraphics[width=\mysize]{pic//show_\name_res\i.png} \\
			\i ~/ residual \vspace{4pt}
		\end{minipage}\hspace{\suojian}
	}
	\\
	\foreach \i in {ZSPan,Ours}
	{
		\begin{minipage}[t]{\mysize*2}
			\centering
			\small
			\includegraphics[width=\mysize]{pic//show_\name_\i.png}\includegraphics[width=\mysize]{pic//show_\name_res\i.png} \\
			\i ~/ residual \vspace{4pt}
		\end{minipage}\hspace{\suojian}
	}
	\caption{Visualization results on WV3 reduced resolution dataset of all compared methods and the proposed PSDip. Both the restored HRMS and the residual image are shown.}
	\label{fig-WV3-reduced}
\end{figure*}

\begin{figure*}[t]
	\newcommand{\mysize}{2.7cm}
	\newcommand{\name}{QB}
	\newcommand{\suojian}{-4pt}
	\centering
	\begin{minipage}[t]{\mysize*2}
		\centering
		\small
		\includegraphics[width=\mysize]{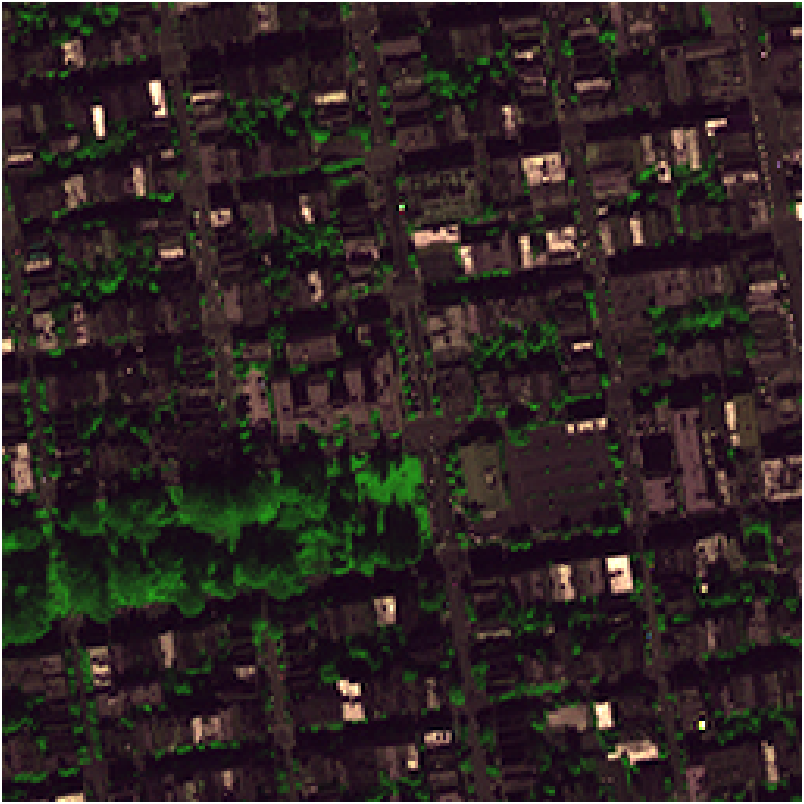}\includegraphics[width=\mysize]{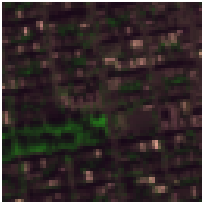} \\
		HRMS / LRMS
	\end{minipage}\hspace{\suojian}
	\foreach \i in {BT-H,BDSD-PC}
	{
		\begin{minipage}[t]{\mysize*2}
			\centering
			\small
			\includegraphics[width=\mysize]{pic//show_\name_\i.png}\includegraphics[width=\mysize]{pic//show_\name_res\i.png} \\
			\i ~/ residual \vspace{4pt}
		\end{minipage}\hspace{\suojian}
	}
	\\
	\foreach \i in {C-GSA, AWLP,GLP-HPM}
	{
		\begin{minipage}[t]{\mysize*2}
			\centering
			\small
			\includegraphics[width=\mysize]{pic//show_\name_\i.png}\includegraphics[width=\mysize]{pic//show_\name_res\i.png} \\
			\i ~/ residual \vspace{4pt}
		\end{minipage}\hspace{\suojian}
	}
	\\
	\foreach \i in {GLP-FS,LDPNet, LRTCFPan}
	{
		\begin{minipage}[t]{\mysize*2}
			\centering
			\small
			\includegraphics[width=\mysize]{pic//show_\name_\i.png}\includegraphics[width=\mysize]{pic//show_\name_res\i.png} \\
			\i ~/ residual \vspace{4pt}
		\end{minipage}\hspace{\suojian}
	}
	\\
	\foreach \i in {ZSPan,Ours}
	{
		\begin{minipage}[t]{\mysize*2}
			\centering
			\small
			\includegraphics[width=\mysize]{pic//show_\name_\i.png}\includegraphics[width=\mysize]{pic//show_\name_res\i.png} \\
			\i ~/ residual \vspace{4pt}
		\end{minipage}\hspace{\suojian}
	}
	\caption{Visualization results on QB reduced resolution dataset of all compared methods and the proposed PSDip. Both the restored HRMS and the residual image are shown.}
	\label{fig-QB-reduced}
\end{figure*}

\begin{figure*}[t]
	\newcommand{\mysize}{2.9cm}
	\newcommand{\name}{QB\_full}
	\newcommand{\suojian}{-5pt}
	\centering
	\foreach \i in {PAN, LRMS, BT-H,BDSD-PC,C-GSA}
	{
		\begin{minipage}[t]{\mysize}
			\centering
			\small
			\includegraphics[width=\mysize]{pic//show_\name_\i.png}\\
			\i  \vspace{4pt}
		\end{minipage}\hspace{\suojian}
	}
	\begin{minipage}[t]{\mysize}
		\centering
		\small
		\includegraphics[width=\mysize]{pic//show_\name_AWLP.png}\\
		AWLP  \vspace{4pt}
	\end{minipage}
	\\
	\foreach \i in {GLP-HPM,GLP-FS,LDPNet,LRTCFPan,ZSPan}
	{
		\begin{minipage}[t]{\mysize}
			\centering
			\small
			\includegraphics[width=\mysize]{pic//show_\name_\i.png}\\
			\i  \vspace{4pt}
		\end{minipage}\hspace{\suojian}
	}
	\begin{minipage}[t]{\mysize}
		\centering
		\small
		\includegraphics[width=\mysize]{pic//show_\name_Ours.png}\\
		Ours  \vspace{4pt}
	\end{minipage}
	\caption{Visualization results on QB full resolution dataset of all compared methods and the proposed PSDip.}
	\label{fig-QB-full}
\end{figure*}

\subsection{Pansharpening results}
In this section, we present the averaged pansharpening results of all compared methods and the proposed method for each dataset in Table \ref{tab-WV2}-\ref{tab-QB-full}. We also provide the standard variance (std) of each index value over the images contained in each dataset. 

Table \ref{tab-WV2}-\ref{tab-QB} shows the results on reduced resolution WV2, WV3 and QB datasets. The proposed PSDip achieves the best performance for all six assessments on the three datasets. This means the proposed method can finely restore the spatial details and also preserve the spectral information. LRTCFPan achieves second-best results for most cases. Table \ref{tab-QB-full} shows the results on the full-resolution QB dataset. Our PSDip also achieves the best performance for the three indexes. This means the proposed method can handle real situations. Besides, the standard variances of the proposed methods are lower compared with other methods in most cases, which means that our method performs quite stable when facing different images. We also present the running time of each method in Table \ref{tab-WV2}-\ref{tab-QB}. We see that the CS- and MRA-based methods require the least time. Unsupervised, VO-based and zero-shot methods take more running time because they need optimization. The proposed PSDip takes more time than LRTCFPan and ZSPan, but it is acceptable.

In Fig. \ref{fig-WV2-reduced}-\ref{fig-QB-full}, we visualize the pansharpening results of all methods on the test datasets. We select the $(8,2,1)$-th bands of WV2/WV3 data and the $(2,4,1)$-th bands of QB data to generate the pseudo-color images. The residual image is composed of the absolute difference between the pseudo-color HRMS and the pseudo-color restored HRMS. We see that the results of BDSD-PC, C-GSA, GLP-HPM, GLP-FS and LDPNet are a little over-smoothed. Particularly, in Fig. \ref{fig-WV3-reduced}, The residual images of LRTCFPan and our PSDip obviously contain less image information. Besides, our method also tends to preserve more texture details than other methods, which reveals the effectiveness of the predicted $\G$ by our method. 

\subsection{\texorpdfstring{Analysis of $\G$ generated by PSDip}{}}
\begin{figure*}[ht]
	\centering
	\includegraphics[width=18cm]{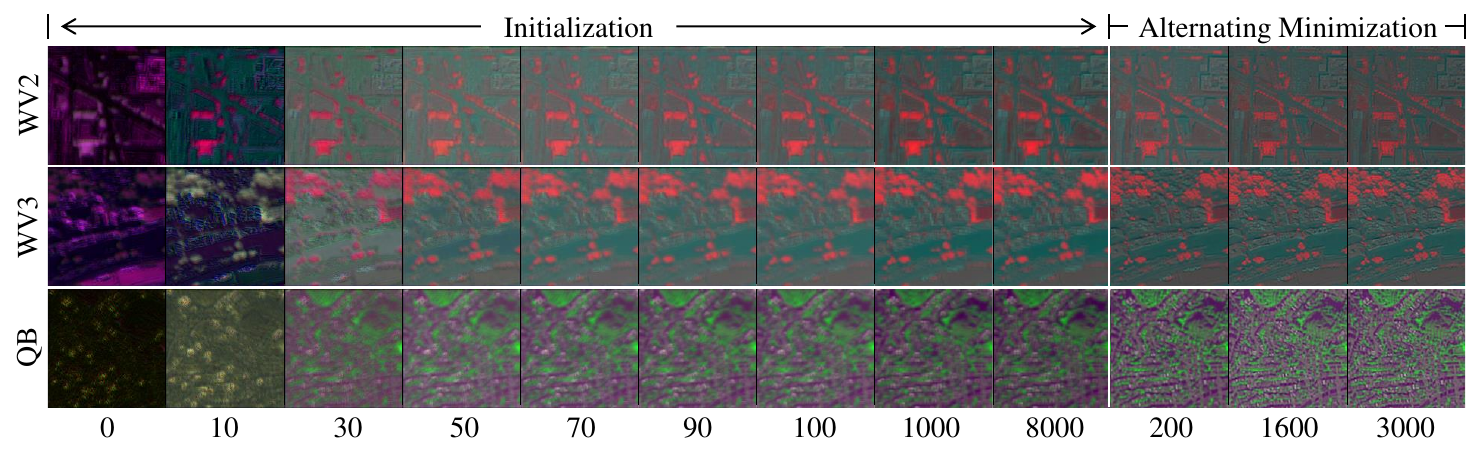}
	\caption{Visualization of $\G$ during the $f_\theta$ initialization process and alternating minimization process for solving (\ref{main-method}).}
	\label{fig-G-process}
\end{figure*}

In the proposed PSDip, we estimate $\G$ by a neural network $f_\theta$. The network is first initialized by (\ref{init-theta}) and then iteratively updated together with $\X$ in (\ref{main-method}). In Fig. \ref{fig-G-process}, we illustrate the intermediate $\G$s generated from both processes to show how $\G$ changes. At the very beginning when $f_\theta$ has not been optimized at all, we see that the generated $\G_0$ already contains a few relevant image information. This observation confirms that the network structure can construct image structures even though the network parameters are randomly set. However, the image information of $\G_0$ does not meet our requirements for pan-sharpening. For the first 100 steps of the initialization process, the generated $\G$ quickly matches the ``spectral" information, which is revealed by the ``color change". In the remaining steps of the initialization process, we see that $\G$ is gradually refined with spatial details as well as spectral information. These refinements make $\G$ look sharper. The final $\G$ of the initialization process looks more like the expected $\G$ but is of course still quite blurry. During the alternating minimization process, we see that there is an apparent quality improvement on $\G$ compared with the initialization process. This is attributed to the formulated optimization problem that can more carefully establish the relationship between HRMS, LRMS and PAN. Besides, $\G$ is also incrementally incorporated with more details during this process. 

\begin{figure}[ht]
	\newcommand{\mysize}{2.62cm}
	\newcommand{\suojian}{-2pt}
	\renewcommand{\arraystretch}{1.15}
	\centering
	\begin{tabular}{c l }
		&\hspace{10pt} $\X_{gt}\oslash\hat{P}$ \hspace{28pt} $f_{\theta_T}(\X_T,P)$ \hspace{18pt} $f_{\theta_T}(\X_{gt},P)$ \\
		\toprule
		\rotatebox{90}{\hspace{0.9cm}WV2} & \hspace{-0.4cm}
		\begin{minipage}[t]{\mysize}
			\centering
			\small
			\includegraphics[width=\mysize]{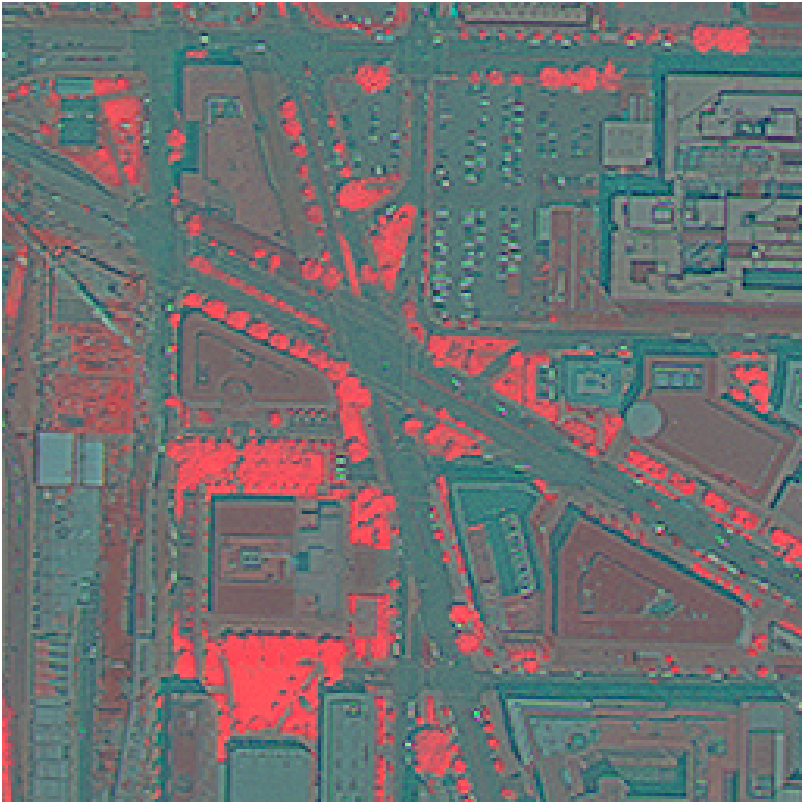} \\
			MSE
		\end{minipage}\hspace{\suojian}
		\begin{minipage}[t]{\mysize}
			\centering
			\small
			\includegraphics[width=\mysize]{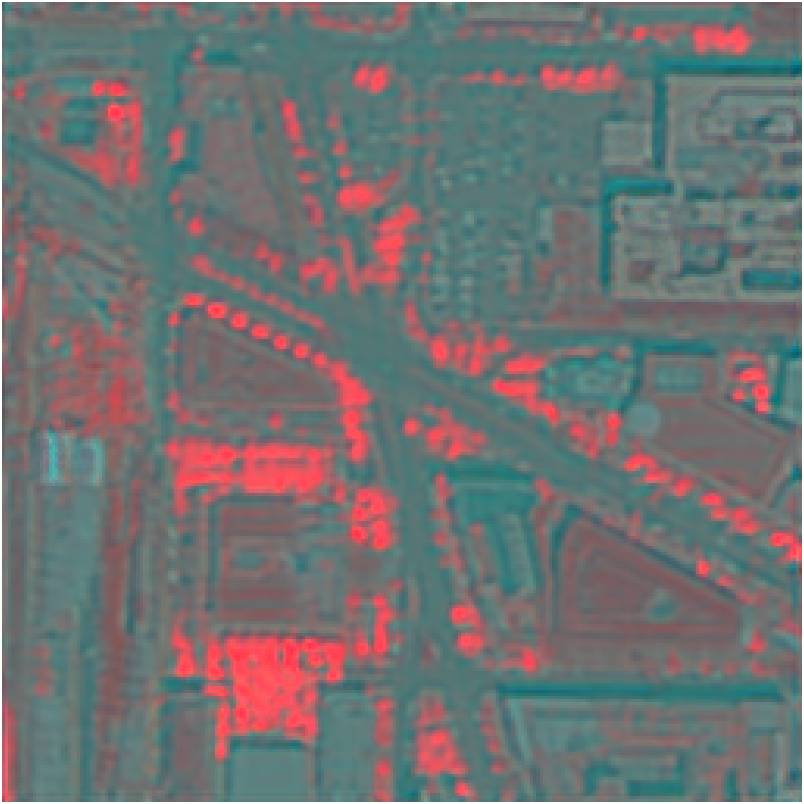} \\
			0.0276
		\end{minipage}\hspace{\suojian}
		\begin{minipage}[t]{\mysize}
			\centering
			\small
			\includegraphics[width=\mysize]{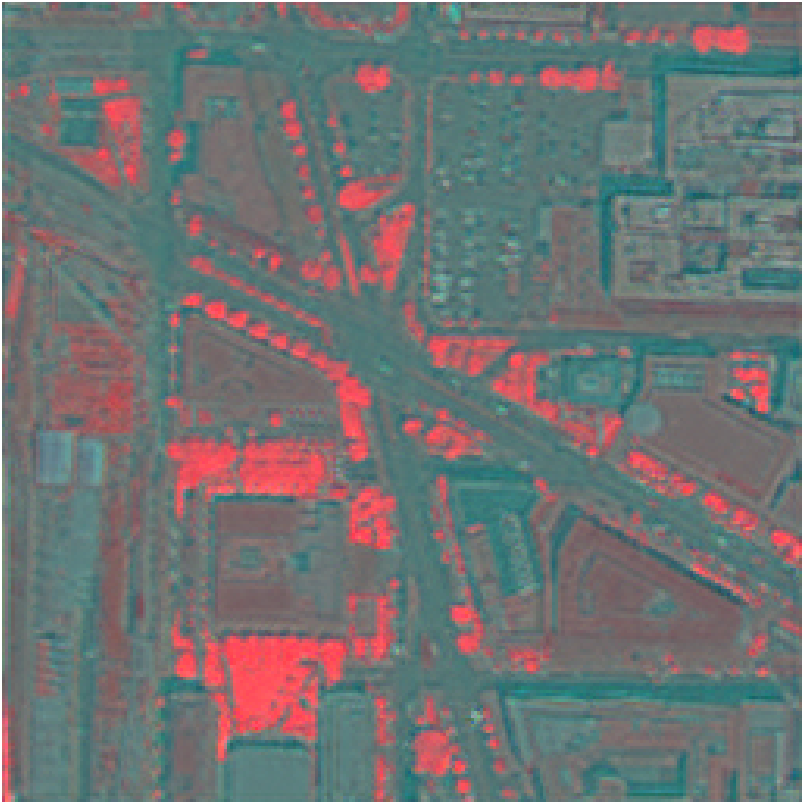} \\
			0.0184
		\end{minipage} \vspace{2pt}\\
		\midrule
		\rotatebox{90}{\hspace{0.9cm}WV3} & \hspace{-0.4cm}
		\begin{minipage}[t]{\mysize}
			\centering
			\small
			\includegraphics[width=\mysize]{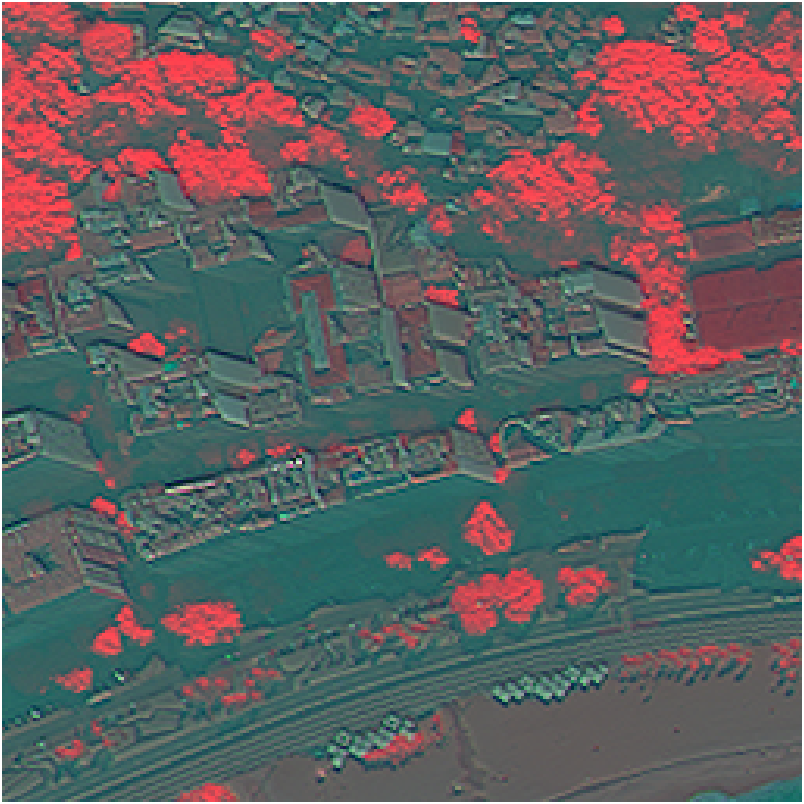} \\
			MSE
		\end{minipage}\hspace{\suojian}
		\begin{minipage}[t]{\mysize}
			\centering
			\small
			\includegraphics[width=\mysize]{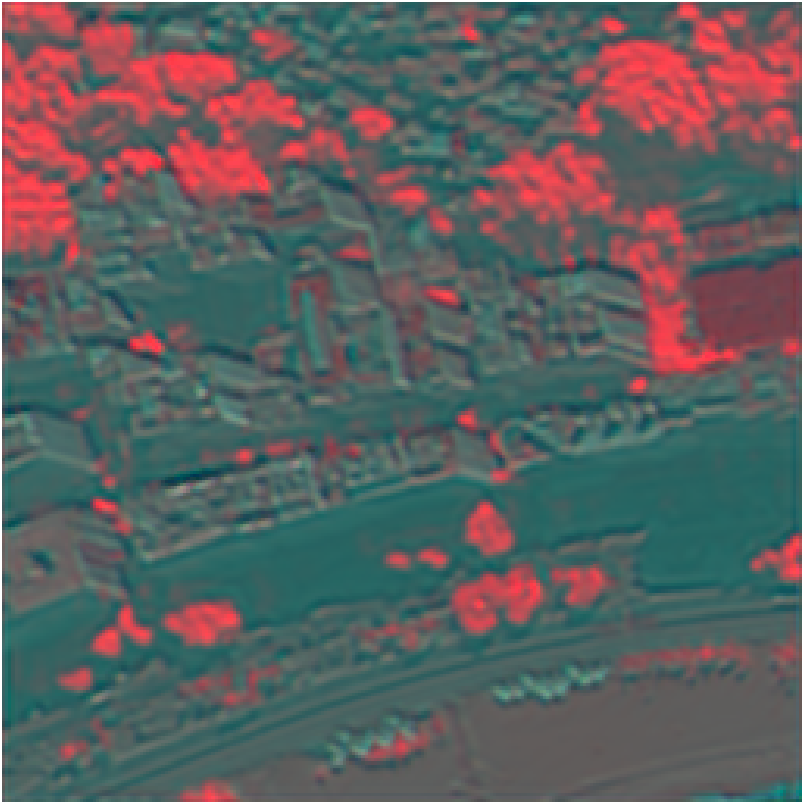} \\
			0.0322
		\end{minipage}\hspace{\suojian}
		\begin{minipage}[t]{\mysize}
			\centering
			\small
			\includegraphics[width=\mysize]{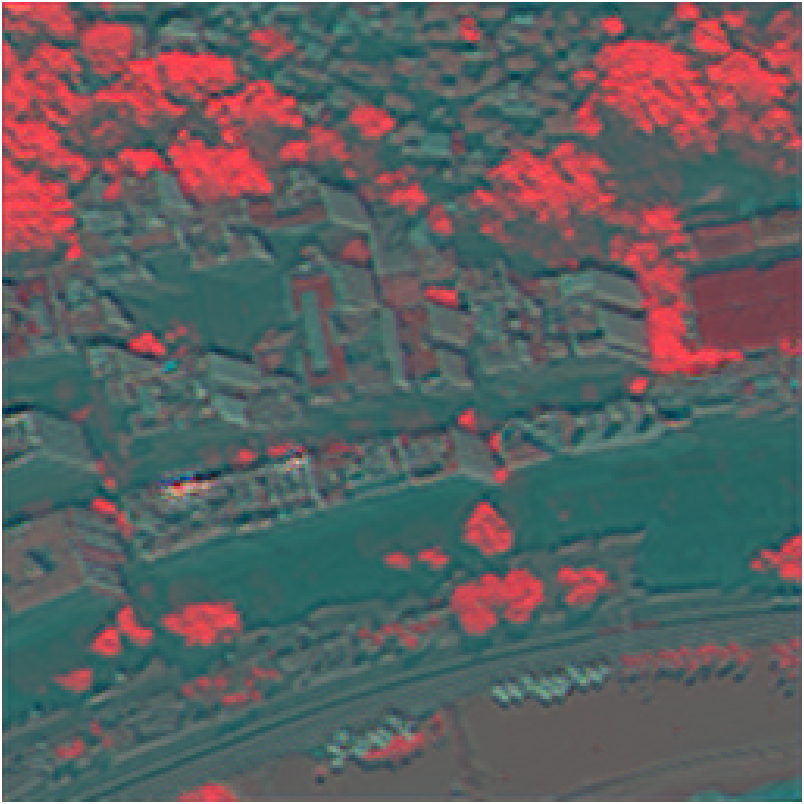} \\
			0.0250
		\end{minipage}\vspace{2pt}\\
		\midrule
		\rotatebox{90}{\hspace{1cm}QB} & \hspace{-0.4cm}
		\begin{minipage}[t]{\mysize}
			\centering
			\small
			\includegraphics[width=\mysize]{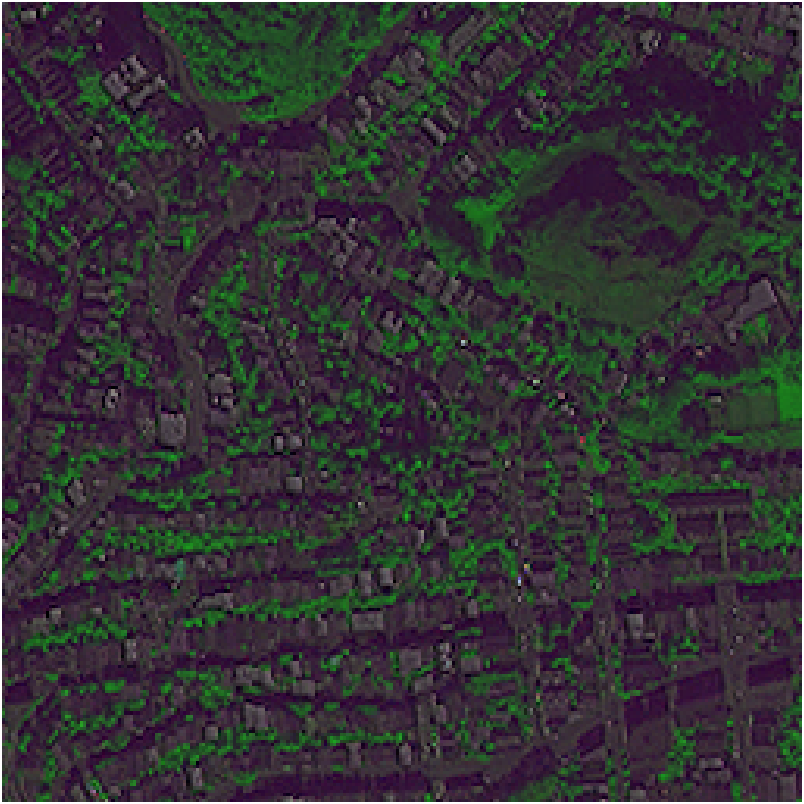} \\
			MSE
		\end{minipage}\hspace{\suojian}
		\begin{minipage}[t]{\mysize}
			\centering
			\small
			\includegraphics[width=\mysize]{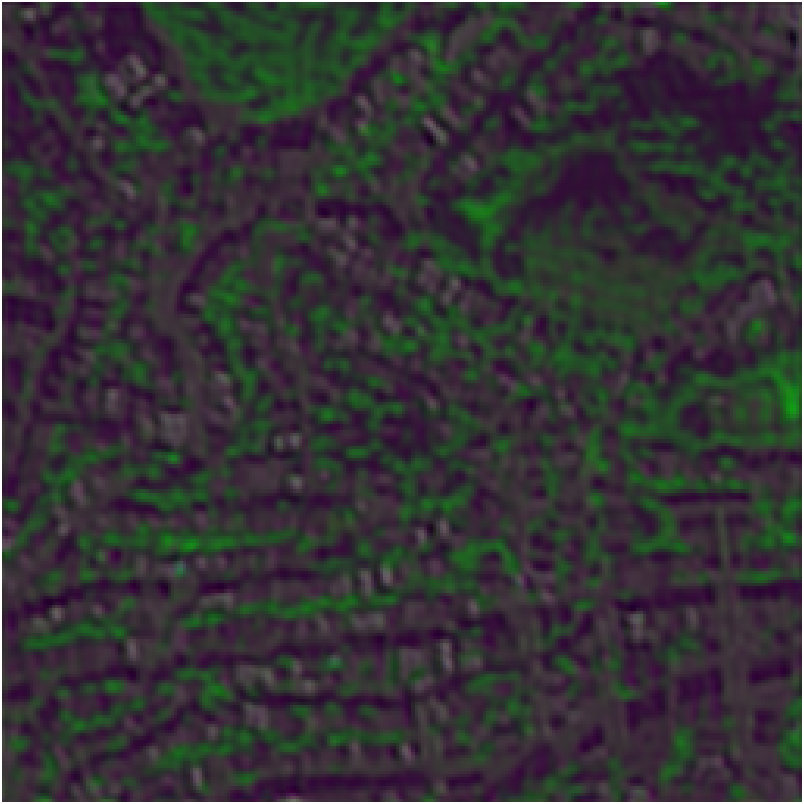} \\
			0.0601
		\end{minipage}\hspace{\suojian}
		\begin{minipage}[t]{\mysize}
			\centering
			\small
			\includegraphics[width=\mysize]{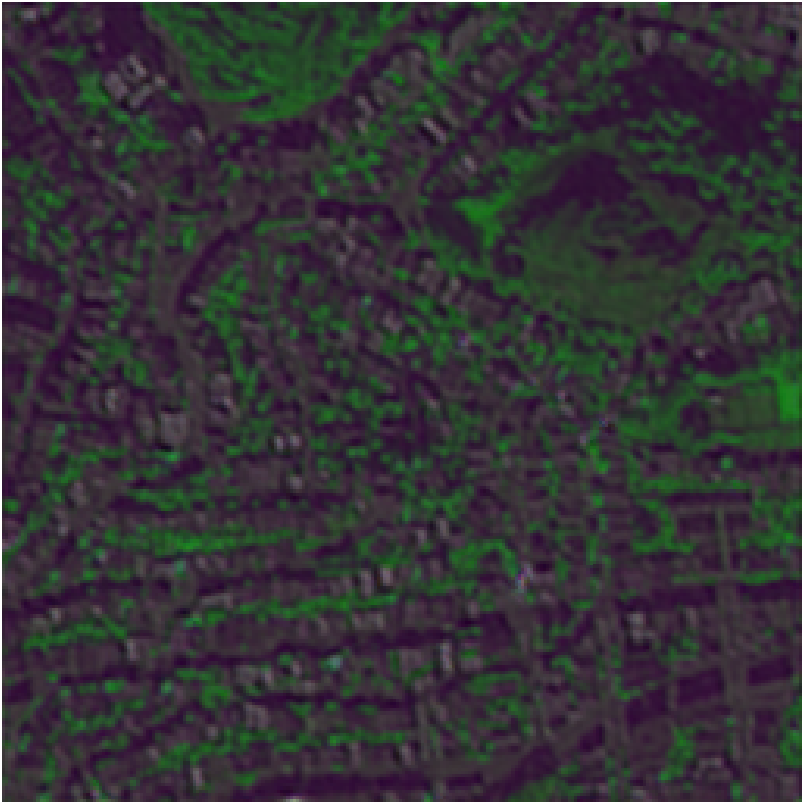} \\
			0.0477
		\end{minipage}
	\end{tabular}
	\caption{Visualization of ground-truth $\G=\X_{gt}\oslash\hat{P}$, the estimated $\G_T=f_{\theta_T}(\X_T,P)$ by our method and $\tilde{\G}=f_{\theta_T}(\X_{gt},P)$ for comparison.}
	\label{fig-dis-G}
\end{figure}

We can calculate three special intermediate $\G$s in the entire above optimization process. When initialization process (\ref{init-theta}) finishes, we get a initialized value of $\G$, i.e., $\G_{init}:=f_{\theta_0^*}(\hat{\Y},P)$. The final estimated $\G$ when the alternating minimization process finishes is denoted as $\G_T := f_{\theta_T}(\X_T, P)$. Besides, we can also put $\X_{gt}$ into the final trained network $f_{\theta_T}$, the derived output is denoted as $\tilde{\G}:=f_{\theta_T}(\X_{gt}, P)$. In Table \ref{tab-dis-G}, we calculate the mean square error (MSE) between the ground-truth $\G=\X_{gt}\oslash\hat{P}$ and $\G_{init}, \G_{T}, \tilde{\G}$, respectively. The first line in Table \ref{tab-dis-G} shows that the initialization (\ref{init-theta}) can produce rough values of $\G$ for the alternating minimization. The second line shows that the final estimated $\G_T$ is more accurate than $\G_{init}$, which reveals the benefits of updating $\G$ in the proposed problem (\ref{main-method}). Compared with the first two lines in Table \ref{tab-dis-G}, the last line gets the smallest error. Specifically, this means that our trained $f_{\theta_T}$ could produce more accurate $\G$ when the inputs are ``ideal". Thus, the trained $f_\theta$ itself by our PSDip should exactly have learned how to predict $\G$ from $\X$ and $P$. Fig. \ref{fig-dis-G} shows visualization of the ground-truth $\G$, our estimated $\G_T$ and $\tilde{\G}$. It can be seen that the coefficient $\G$ can be finely estimated by our PSDip. 

\begin{table}[ht]
	\renewcommand{\arraystretch}{1.29}
	\newcommand{\mysize}{1cm}
	\fontsize{9}{10}\selectfont
	\caption{Mean square error (MSE) between groundtruth $\G=\X_{gt}\oslash\hat{P}$ and three intermediate $\G$s, i.e., $\G_{init}$, $\G_T$ and $\tilde{\G}$. The best results are in \textbf{bold}.}
	\label{tab-dis-G}
	\centering
	\begin{tabular}{ L{3cm} | M{\mysize} M{\mysize} M{\mysize}}
		\Xhline{1pt}
		& WV2 & WV3 & QB \\
		\hline
		$\mathrm{MSE}(\X_{gt}\oslash\hat{P}, \G_{init})$ & 0.0344 & 0.0298 & 0.0667 \\
		$\mathrm{MSE}(\X_{gt}\oslash\hat{P}, \G_T)$ & 0.0221 & 0.0161 & 0.0409\\
		$\mathrm{MSE}(\X_{gt}\oslash\hat{P}, \tilde{\G})$ & \tb{0.0148} & \tb{0.0124} & \tb{0.0322}\\
		\Xhline{1pt}
	\end{tabular}
\end{table}

\subsection{Parameter analysis}\label{sec-parameter-setting}
\begin{figure}[ht]
	\newcommand{\mysize}{8.5cm}
	\newcommand{\suojian}{-5pt}
	\centering
	\begin{minipage}[t]{\mysize}
		\centering
		\small
		\includegraphics[width=\mysize]{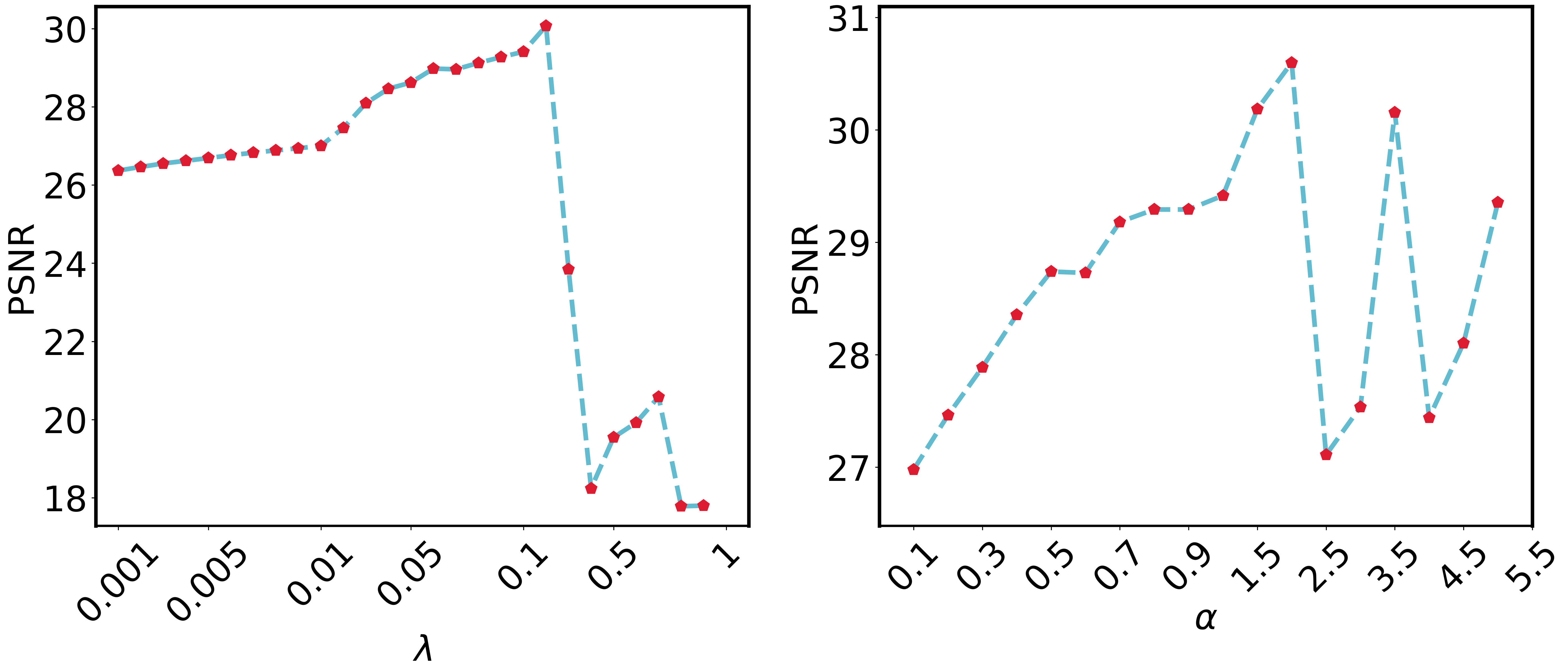}\\
	\end{minipage}
	\caption{Analysis of parameters $\lambda$ and $\alpha$ using five images from WV3 validation dataset.}
	\label{fig-lam-alpha}
\end{figure}

There are mainly two parameters to be set. One is the only trade-off parameter $\lambda$ in the proposed problem (\ref{main-method}). The other one is the step size $\alpha$ for updating $\X$ in (\ref{update-X}). Five images from the WV3 validation dataset are used to analyze the two parameters. We present the results in Fig. \ref{fig-lam-alpha}. When $\lambda$ is less than 0.2, its growing value enables the generated coefficient $\G = f_{\theta}(\X,P)$ to contribute more to the restored HRMS and the PSNR value then gradually rises. However, when $\lambda$ keeps growing exceeding 0.2, the spectral fidelity term $L_p$ occupies too much proportion in the optimization objective. The training of $f_\theta$ tends to be unstable because both $\X$ and $\theta$ are undetermined in $L_p$. Thus, we set $\lambda=0.1$ for all experiments to have a good balance between $L_y$ and $L_p$. For the step size $\alpha$, when its value is greater than 2, we see that the alternating minimization process also becomes unstable. Thus, we set $\alpha=2$ for all the experiments.

\begin{figure}[ht]
	\newcommand{\mysize}{4.4cm}
	\newcommand{\suojian}{-7pt}
	\centering
	\begin{minipage}[t]{\mysize}
		\centering
		\small
		\includegraphics[width=\mysize]{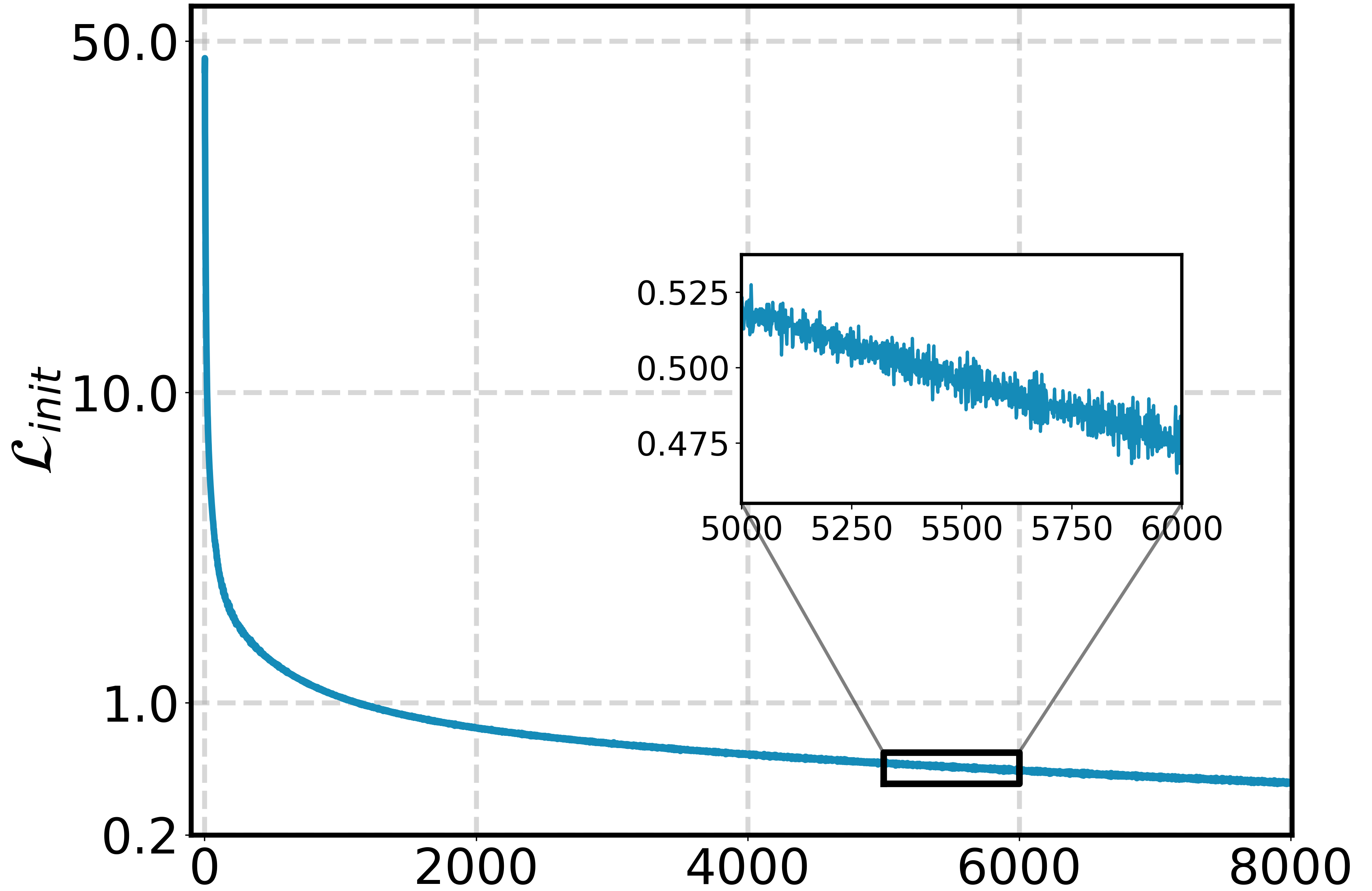}\\
		Initialization
	\end{minipage}
	\hspace{\suojian}
	\begin{minipage}[t]{\mysize}
		\centering
		\small
		\includegraphics[width=\mysize]{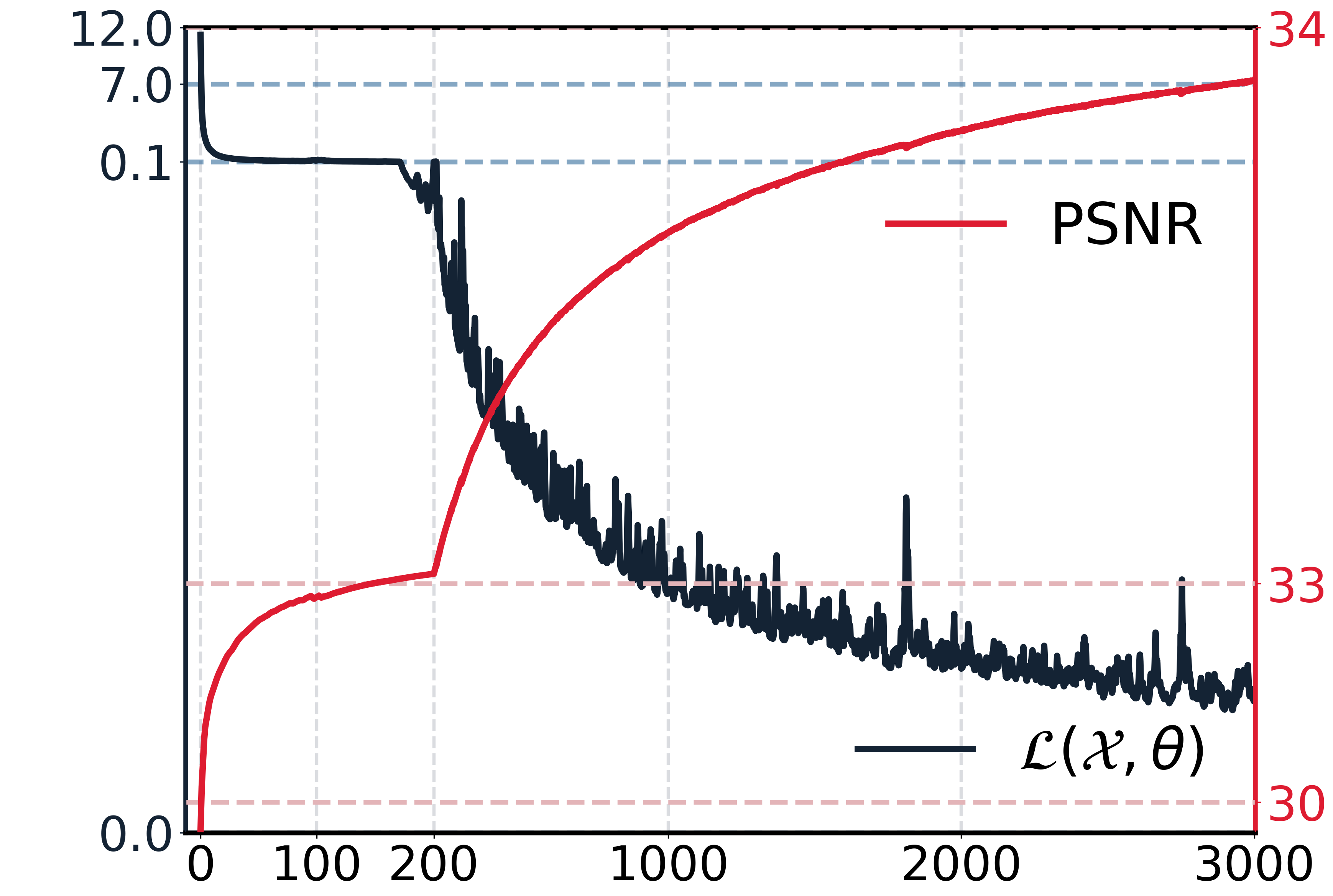}\\
		Alternating minimization
	\end{minipage}
	\caption{Trends of losses or PSNR in the initialization and alternating minimization process on QB test dataset.}
	\label{fig-curve-loss}
\end{figure}

In Fig. \ref{fig-curve-loss}, we plot the loss and PSNR trends in the initialization and alternating minimization process on the QB test dataset. Let $\la_{init}$ denote the loss function in (\ref{init-theta}) for initializing $f_\theta$:
\begin{align}
	\la_{init}:=\|\hat{\Y} - f_\theta(\hat{\Y}, P)\odot(\hat{P}\otimes K)\|.
\end{align}
The left image plots the averaged $\la_{init}$ on 20 images contained in the dataset. We see that the loss quickly drops at the first hundreds of steps, which is in consistent with the observation in Fig \ref{fig-G-process}. The right image shows the trends of main objective $\la(\X, P)$ in (\ref{main-method}) and the corresponding PSNR values. The objective also quickly drops at the first dozens of steps and then gently converges. The PSNR value keeps gently growing when the iteration step is greater than 1000, in which stage we think the restored HRMS is gradually refined and added with more details. For all experiments, the initialization process takes 8000 optimization steps and the alternating minimization process takes 3000 steps.

\subsection{\texorpdfstring{Ablation studies of $f_\theta$}{}}
In this section, we conduct several ablation studies about settings of the network $f_\theta$ of our PSDip. The experiments use five images from the WV3 validation dataset. Specifically, the following three cases are considered.

\textit{case 1}: Like most previous DIP methods, $f_\theta$ takes random noise $z\sim\mathcal{N}(0,1)$ as input, i.e., $\G = f_\theta(z)$.

\textit{case 2}: $f_\theta$ still takes $\X$ and $P$ as inputs like PSDip. However, during the alternating minimization process, only $\X$ is optimized and $\theta$ is fixed as $\theta_0^*$.

\textit{case 3}: $f_\theta$ still takes $\X$ and $P$ as inputs like PSDip. However, it is not initialized by (\ref{init-theta}) but randomly initialized for the alternating minimization process.

The results are presented in Table \ref{tab-dis-f}. Comparing case 1 and our PSDip, we can see that our inputs $(\X, P)$ could provide more information for generating $\G$ than random input $z$ and thus our PSDip has better performance. Results of case 2 show that updating $\theta$, which also means updating $\G$, for the proposed pansharpening problem (\ref{main-method}) is useful, although the initialization process could produce a relatively ``good" $\G$. Results of case 3 show that directly starting optimizing $f_\theta$ in the main objective (\ref{main-method}) brings more uncertainty and does not produce satisfying results. Furthermore, in Fig. \ref{fig-curve-noinit}, we present the losses and PSNR values of case 3 and PSDip during the alternating minimization process. It shows that with a good initial value of $\theta$, the optimization process could be much more stable and also helps our model to get better performance.
\begin{table}[ht]
	\renewcommand{\arraystretch}{1.29}
	\newcommand{\mysize}{1.4cm}
	\fontsize{9}{10}\selectfont
	\caption{Results of different settings of $f_\theta$ on WV3 validation dataset. The best results are in \tb{bold}.}
	\label{tab-dis-f}
	\centering
	\begin{tabular}{ L{0.8cm} | M{\mysize} M{\mysize-0.2cm} M{\mysize} | M{0.6cm} M{0.75cm}}
		\Xhline{1pt}
		& initialize $f_\theta$ by (\ref{init-theta}) & $f_\theta$ takes random input & fix $\theta$ as $\theta_0^*$ in (\ref{main-method}) & PSNR & SSIM\\
		\hline 
		case 1  & \dui & \dui & \bdui & 28.95 & 0.8772\\
		case 2  & \dui & \bdui & \dui & 28.55 & 0.8710 \\
		case 3  & \bdui & \bdui & - & 26.92 & 0.7775 \\
		PSDip & \dui & \bdui & \bdui & \tb{30.51} & \tb{0.9156} \\
		\Xhline{1pt}
	\end{tabular}
\end{table}

\begin{figure}[ht]
	\newcommand{\mysize}{4.3cm}
	\newcommand{\suojian}{-6pt}
	\centering
	\begin{minipage}[t]{\mysize}
		\centering
		\small
		\includegraphics[width=\mysize]{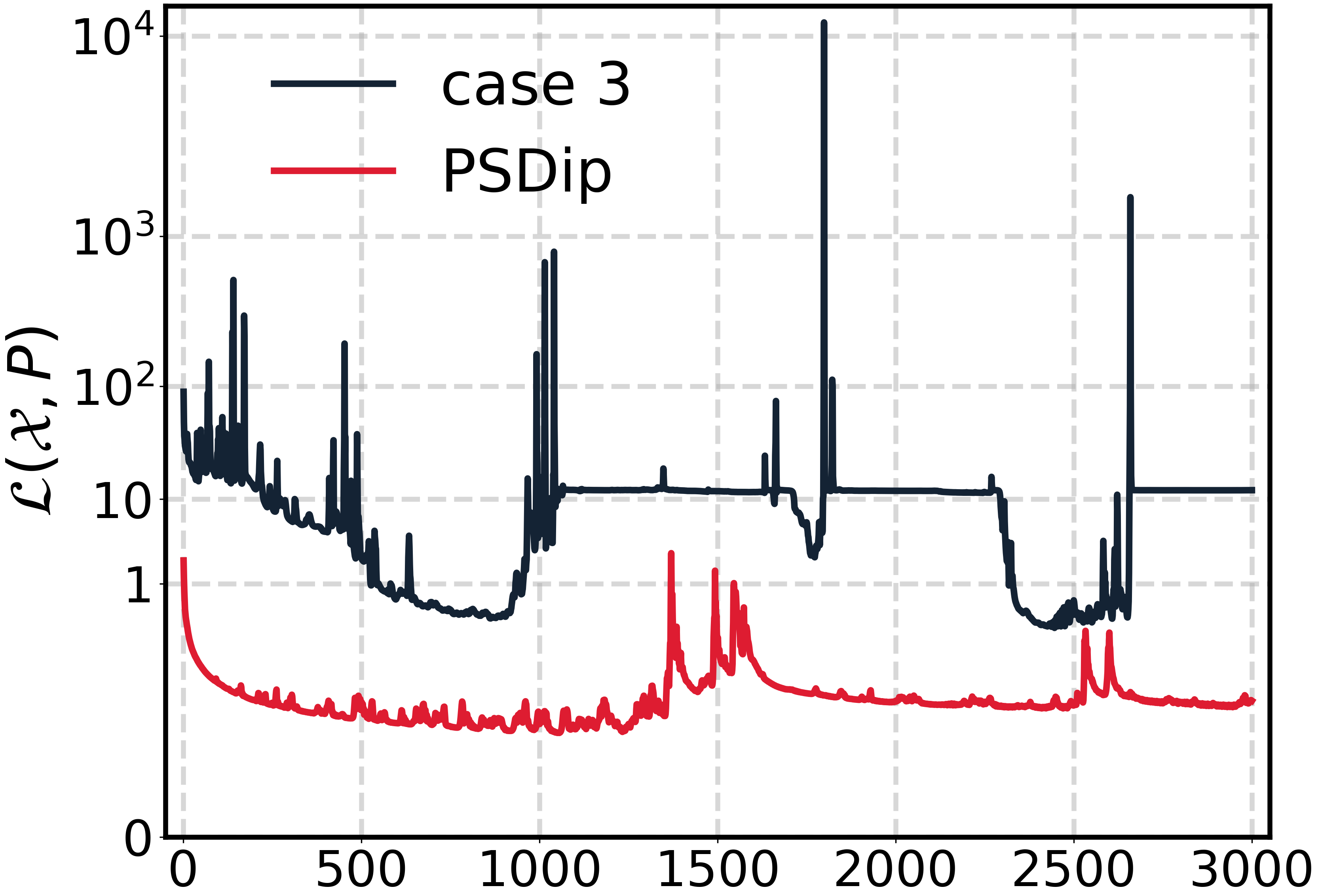}\\
		Initialization
	\end{minipage}
	\hspace{\suojian}
	\begin{minipage}[t]{\mysize}
		\centering
		\small
		\includegraphics[width=\mysize]{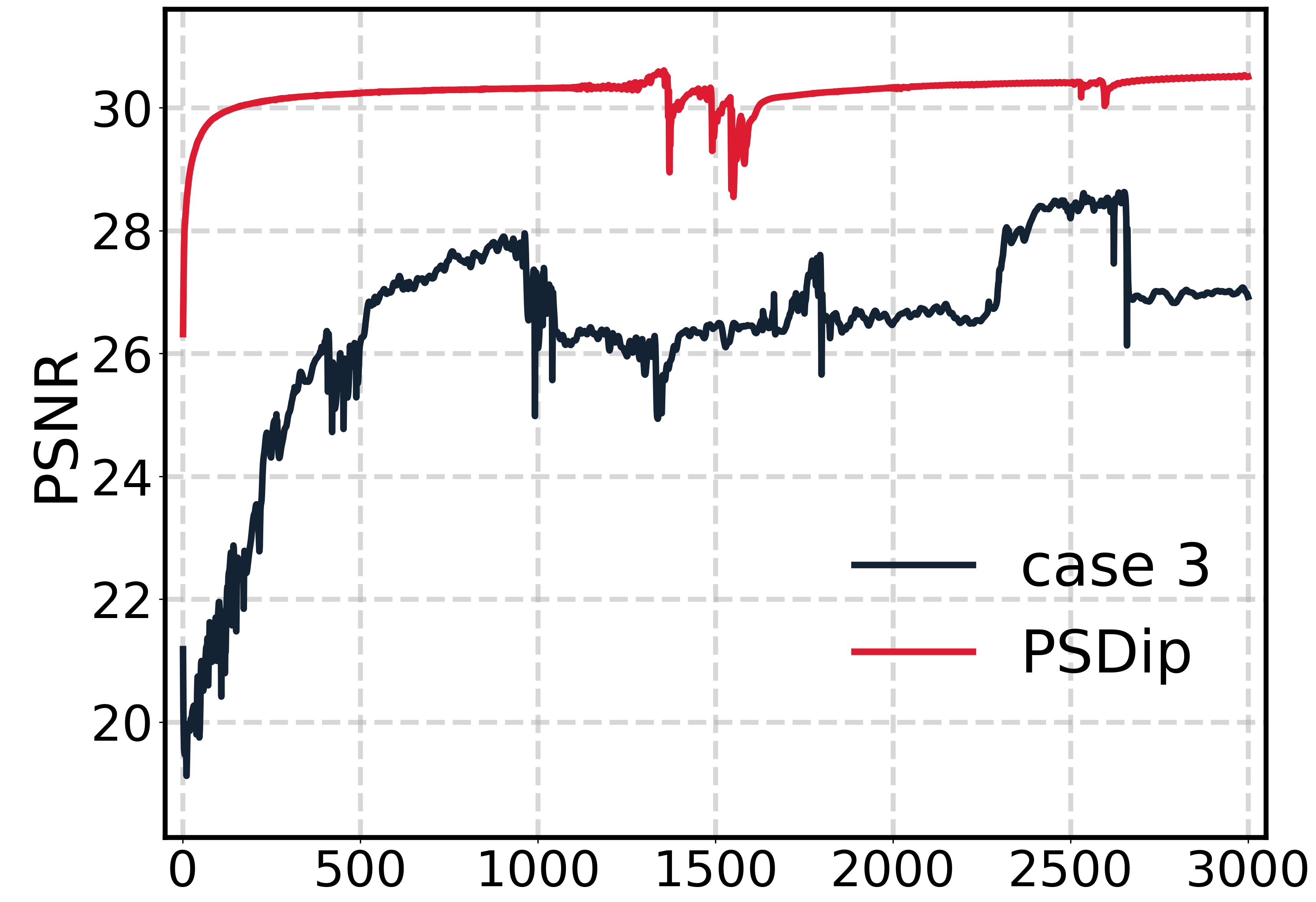}\\
		Alternating minimization
	\end{minipage}
	\caption{Effectiveness of \eqref{init-theta}: trends of losses and PSNR in the alternating minimization process on WV3 validation dataset. Case3 does not initialize $f_\theta$ by $\min_\theta\mathcal{L}_{init}$ but uses a randomly initialized $f_\theta$ for alternating minimization.}
	\label{fig-curve-noinit}
\end{figure} 

\subsection{\texorpdfstring{Analysis of updating $\X$}{}}\label{sec-ana-X}
In Sec. \ref{sec-solve-the-proposed-method}, we update $\X$ by solving a simplified $\X$-subproblem $\la_\X$. For convenience, let ${\X_f}_t$ denote the network input $\X$ in the $t$-th step. The simplified $\X$-subproblem fix ${\X_f}_t$ as $\X_{t-1}$ in the $t$-th step. Otherwise, ${\X_f}_t$ can also be set as the to-be-updated variable $\X$. The corresponding objective of this $\X$-subproblem is
\begin{align}
	&\la'_\X(\X, \theta_{t-1}) := \nonumber \\
	& \left\|\Y \!-\! (\X\otimes K)\downarrow_r \right\|_F^2\! +\! \lambda \left\|\X\! -\! f_{\theta_{t-1}}(\X, P)\odot\hat{P}  \right\|_F^2.
\end{align}
In Table \ref{tab-ana-X}, we present the performance by these two settings of ${\X_f}_t$. It shows that updating $\X$ using $\la'_\X$ performance worse than the proposed PSDip. This implies that the gradient of network $f_\theta$ with respect to $\X$ does not necessarily help to better optimize $\X$ in problem (\ref{main-method}).

\begin{table}[ht]
	\renewcommand{\arraystretch}{1.29}
	\newcommand{\mysize}{1.65cm}
	\fontsize{9}{10}\selectfont
	\caption{``PSNR/SSIM" results between two setting of ${\X_f}_t$ on reduced resolution WV2, WV3, QB datasets. In the first case, ${\X_f}$ is set as variable $\X$. The second case (i.e. PSDip) fixes ${\X_f}_t$ as $\X_{t-1}$.}
	\label{tab-ana-X}
	\centering
	\begin{tabular}{ M{2.1cm} | M{\mysize} M{\mysize} M{\mysize}}
		\Xhline{1pt}
		& WV2 & WV3 & QB \\
		\hline
		not fix ${\X_f}_t$ & 27.34/0.7329 & 32.64/0.8874 & 31.60/0.7998\\
		fix ${\X_f}_t$ (PSDip) & \tb{31.17}/\tb{0.8593} & \tb{34.43}/\tb{0.9205} & \tb{34.10}/\tb{0.8925} \\ 
		\Xhline{1pt}
	\end{tabular}
\end{table}

\subsection{Comparing $\G$ with HRMS}
\begin{figure}[ht]
	\centering
	\newcommand{\mysize}{2.77cm}
	\begin{minipage}[t]{\mysize}
		\centering
		\includegraphics[width=\mysize]{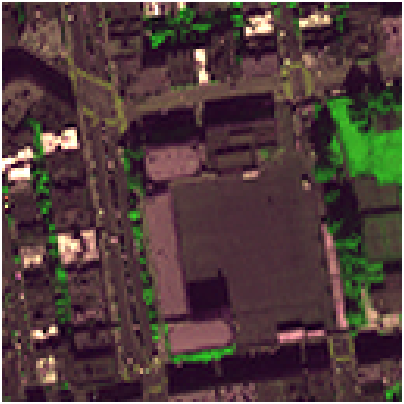} \\
		Take $\Y$ as HRMS
	\end{minipage}
	\begin{minipage}[t]{\mysize}
		\centering
		\includegraphics[width=\mysize]{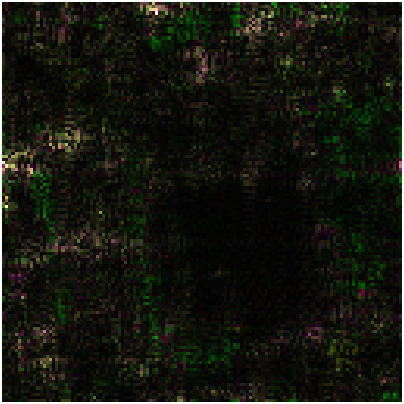} \\
		High frequency
	\end{minipage}
	\begin{minipage}[t]{\mysize}
		\centering
		\includegraphics[width=\mysize]{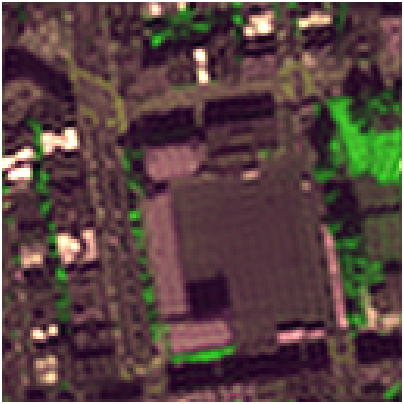} \\
		Low frequency
	\end{minipage} \vspace{4pt}\\
	
	\begin{minipage}[t]{\mysize}
		\centering
		\includegraphics[width=\mysize]{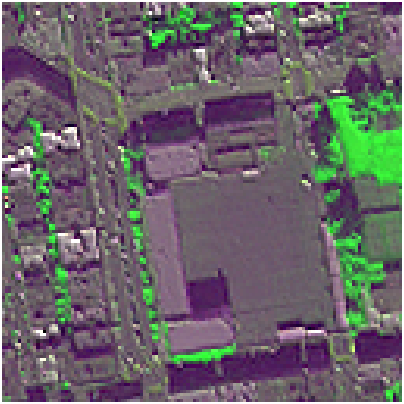} \\
		Coefficient $\G_l$
	\end{minipage}
	\begin{minipage}[t]{\mysize}
		\centering
		\includegraphics[width=\mysize]{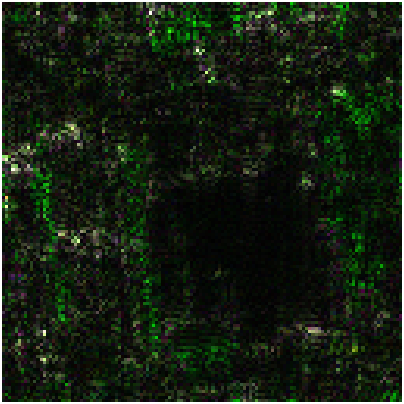} \\
		High frequency
	\end{minipage}
	\begin{minipage}[t]{\mysize}
		\centering
		\includegraphics[width=\mysize]{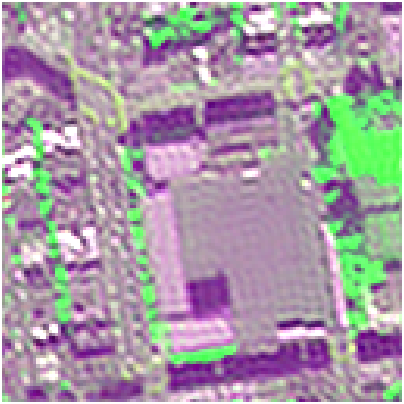} \\
		Low frequency
	\end{minipage}
	\caption{Compare the high- and low-frequency parts of $\G$ and HRMS.}
	\label{fig-consistent}
\end{figure}

According to Wald Protocol \cite{Wald1, Wald2}, we generate degraded LRMS $\Y_l$ and PAN $P_l$ from original LRMS $\Y$ and PAN $P$ by blurring and downsampling. The tools we use are provided by \cite{A-benchmarking-protocol-for-pansharpening-Dataset-preprocessing-and-quality-assessment}.The original LRMS $\Y$ is treated as HRMS. Then $\G_l$ is calculated by $\G_l = \Y\oslash\hat{P}_l$, where $\hat{P}_l = \hat{P}_l' + 0.01$ and $\hat{P}_l'$ is generated by histogram-matching from $P_l$ and $\Y_l$. We use Fourier transform to separate the low and high frequency part from a image. Since $\G$ and HRMS have different physical meanings, we only visually compare them. Fig. \ref{fig-consistent} shows an example from full resolution QB dataset. Both the reduced LRMS  $\Y_l$ and $\G_l$ are rescaled to [0,1]. We see in this experiment the high-frequency maps and low-frequency maps between $\G_l$ and $\Y$ have many similarities. But it seems that the $\G_l$ contains more high-frequency components.

\subsection{\texorpdfstring{Discussions of $R(\G)$}{}}\label{sec-disG}
The proposed PSDip uses a deep neural network to generate the coefficient $\G$, where the network structure is seen as regularization on $\G$. Thus, different networks can be seen as different $R(\G)$s. In this section, we consider two other network as the backbone of our $f_\theta$. They are LAGnet \cite{LAGConv} and PNN \cite{Pansharpening-by-convolutional-neural-networks}. Quantatitive results on reduced resolution WV2, WV3 and QB datasets are presented in Table \ref{tab-multi}. It first shows that changing the backbone network could also produce satisfying fusion results. This implies the effectiveness of deep image prior for the coefficient. Besides, we see that the performance of the three backbone networks has distinctions because they contains different image priors. For WV2 and QB datasets, LAGnet has overall better results. However, it requires much more running time.

\begin{table*}[t]
	\renewcommand{\arraystretch}{1.29}
	\newcommand{\mysize}{1.9cm}
	\fontsize{9}{10}\selectfont
	\caption{Results of different backbone networks on the reduced resolution WV2, WV3 and QB dataset. The backbone networks are LAGnet, PNN, and PanNet. ``T" means the running time. The best results are in \textbf{bold}}.
	\label{tab-multi}
	\centering
	\begin{tabular}{L{1.7cm} | M{\mysize+0.2cm} M{\mysize} M{\mysize} M{\mysize+0.2cm} M{\mysize+0.2cm} M{\mysize} | M{0.6cm}}
		\Xhline{1pt}
		\multicolumn{8}{c}{WV2} \\
		\hline
		backbone & PSNR$\uparrow\pm$ std & SSIM$\uparrow\pm$ std & Q$2^n$$\uparrow\pm$ std &         SAM$\downarrow\pm$ std & ERGAS$\downarrow\pm$ std & SCC$\uparrow\pm$ std & \text{T (s)}  \\ 
		\hline 
		LAGnet\cite{LAGConv} & \tb{31.278} $\pm$ 2.110 & \tb{0.876} $\pm$ \tb{0.025} & \tb{0.471} $\pm$ \tb{0.078} & \tb{5.424} $\pm$ \tb{0.697} & 4.363 $\pm$ 1.724 & 0.892 $\pm$ 0.094 & 979.4 \\
		PNN\cite{Pansharpening-by-convolutional-neural-networks} & 30.823 $\pm$ 1.963 & 0.849 $\pm$ 0.029 & 0.382 $\pm$ 0.092 & 5.529 $\pm$ 0.718 & 4.045 $\pm$ 0.588 & 0.924 $\pm$ \tb{0.007} & 97.5 \\
		PanNet\cite{PanNet-A-deep-network-architecture-for-pan-sharpening} & 31.172 $\pm$ \tb{1.945} & 0.859 $\pm$ 0.028 & 0.417 $\pm$ 0.095 & 5.527 $\pm$ 0.710 & \tb{3.901} $\pm$ \tb{0.536} & \tb{0.931} $\pm$ 0.008 & 279.0 \\ 
		\Xhline{1pt}
		\multicolumn{8}{c}{WV3} \\
		\hline
		LAGnet\cite{LAGConv} & 34.079 $\pm$ 2.525 & \tb{0.930} $\pm$ \tb{0.018} & 0.628 $\pm$ \tb{0.116} & \tb{4.314} $\pm$ \tb{1.255} & 5.322 $\pm$ 3.797 & 0.867 $\pm$ 0.146 & 980.6 \\
		PNN\cite{Pansharpening-by-convolutional-neural-networks} & 33.756 $\pm$ 2.673 & 0.906 $\pm$ 0.023 & 0.546 $\pm$ 0.138 & 4.690 $\pm$ 1.369 & 4.154 $\pm$ 1.244 & 0.935 $\pm$ 0.018 & 97.7 \\
		PanNet\cite{PanNet-A-deep-network-architecture-for-pan-sharpening} & \tb{34.430} $\pm$ \tb{2.518} & 0.921 $\pm$ 0.019 & \tb{0.600} $\pm$ 0.129 & 4.460 $\pm$ 1.265 & \tb{3.821} $\pm$ \tb{1.155} & \tb{0.948} $\pm$ \tb{0.016} & 269.6 \\
		\Xhline{1pt}
		\multicolumn{8}{c}{QB} \\
		\hline
		LAGnet\cite{LAGConv} & \tb{34.441} $\pm$ \tb{2.779} & \tb{0.899} $\pm$ \tb{0.025} & \tb{0.794} $\pm$ \tb{0.089} & 6.748 $\pm$ \tb{1.509} & \tb{6.070} $\pm$ \tb{0.551} & \tb{0.941} $\pm$ \tb{0.011} & 945.0 \\
		PNN\cite{Pansharpening-by-convolutional-neural-networks} & 33.728 $\pm$ 2.896 & 0.880 $\pm$ 0.032 & 0.743 $\pm$ 0.111 & 7.045 $\pm$ 1.591 & 6.596 $\pm$ 0.678 & 0.930 $\pm$ 0.013 & 69.3 \\
		PanNet\cite{PanNet-A-deep-network-architecture-for-pan-sharpening} & 34.102 $\pm$ 2.825 & 0.892 $\pm$ 0.028 & 0.760 $\pm$ 0.103 & \tb{6.744} $\pm$ 1.522 & 6.275 $\pm$ 0.615 & 0.940 $\pm$ 0.012 & 209.3 \\ 
		\Xhline{1pt}
	\end{tabular}
\end{table*}

Next, we also consider replacing the deep image prior with other regularization forms in the pansharpening model (\ref{our-model}). As mentioned in Sec. \ref{sec-proposed-method}, $\G$ contains image structures. The positions of high-frequency and low-frequency parts of $\G$ are consistent with HRMS. Thus, here, we consider regularizing $\G$ by widely-used Total Variation (TV). By specifying $R(\cdot)$ in the model (\ref{our-model}) as the TV norm, we derive the following multispectral pansharpening model:
\begin{align}\label{TV-model}
	\min_{\X,\G}~\left\|\Y - (\X\otimes K)\downarrow_r \right\|_F^2 + \lambda_1\left\|\X - \G\odot\hat{P}\right\|_F^2 + \lambda_2 \|\G\|_{\mathrm{TV}},
\end{align}
where the specific form of TV norm is
\begin{align}
	\|\G\|_{\mathrm{TV}} := \sum_{k=1}^{S}& \left[ \sum_{i=1}^{H-1}\sum_{j=1}^{W}|\G(i+1,j,k) - \G(i,j,k)| \right. \nonumber \\ 
	& \left. + \sum_{i=1}^{H}\sum_{j=1}^{W-1}|\G(i,j+1,k) - \G(i,j,k)| \right]
\end{align}
We call problem (\ref{TV-model}) PS-TV. In this model, $\G$ itself is a variable. Both $\X$ and $\G$ can also be conveniently optimized by alternating minimization. In each step, $\X$ and $\G$ are updated about their respective subproblem. Like PSDip, we also used one step of gradient descent to update $\X$ as well as $\G$ and derive the updating forms like (\ref{update-X}). The initial value of $\G$ is set as zero.

In Table \ref{tab-tv}, we present the results of PS-TV on reduced resolution WV2, WV3 and QB datasets. First, we can clearly see the effectiveness of PS-TV. This further proves that with appropriate regularization on the coefficient $\G$ by representation $\X\approx\G\odot\hat{P}$, the proposed model (\ref{our-model}) is able to solve multispectral pansharpening problem and acquire satisfying results. Second, in most cases, the PSDip performs better and is more stable than PS-TV. Overall, we can see the benefits of regularizing $\G$ by neural network. Specifically, we think the network parameters can be dynamically adjusted during the optimization process to explore the unknown relationship between HRMS and PAN except for the constraints induced by network structure. Except for TV regularization and deep image prior, we believe more regularizations on $\G$ are hopefully to be investigated. 

\begin{table*}[t]
	\renewcommand{\arraystretch}{1.29}
	\newcommand{\mysize}{1.9cm}
	\fontsize{9}{10}\selectfont
	\caption{Results of PS-TV and PSDip on the reduced resolution WV2, WV3 and QB datasets. The best results are in \textbf{bold}.}
	\label{tab-tv}
	\centering
	\begin{tabular}{M{1cm} | M{1cm} | M{\mysize+0.2cm} M{\mysize} M{\mysize} M{\mysize+0.2cm} M{\mysize+0.2cm} M{\mysize}}
		\Xhline{1pt}
		Dataset & Methods & PSNR$\uparrow\pm$ std & SSIM$\uparrow\pm$ std & Q2$^n$$\uparrow\pm$ std &         SAM$\downarrow\pm$ std & ERGAS$\downarrow\pm$ std & SCC$\uparrow\pm$ std  \\ 
		\hline
		\multirow{2}{*}{WV2} & PS-TV & 31.10 $\pm$ 2.02 & 0.854 $\pm$ 0.029 & 0.412 $\pm$ \tb{0.090} & \tb{5.518} $\pm$ \tb{0.692} & 3.939 $\pm$ \tb{0.509} & 0.922 $\pm$ 0.009 \\  
		& PSDip & \tb{31.17} $\pm$ \tb{1.94} & \tb{0.859} $\pm$ \tb{0.028} & \tb{0.417} $\pm$ 0.095 & 5.527 $\pm$ 0.710 & \tb{3.901} $\pm$ 0.536 & \tb{0.931} $\pm$ \tb{0.008}  \\ 
		\hline
		\multirow{2}{*}{WV3} & PS-TV & 33.60 $\pm$ 2.68 & 0.900 $\pm$ 0.022 & 0.551 $\pm$ 0.141 & 4.878 $\pm$ 1.354 & 4.263 $\pm$ 1.258 & 0.928 $\pm$ 0.019 \\
		& PSDip & \tb{34.43} $\pm$ \tb{2.52} & \tb{0.921} $\pm$ \tb{0.019} & \tb{0.600} $\pm$ \tb{0.129} & \tb{4.460} $\pm$ \tb{1.265} & \tb{3.821} $\pm$ \tb{1.155} & \tb{0.948} $\pm$ \tb{0.016} \\
		\hline
		\multirow{2}{*}{QB} & PS-TV & 33.34 $\pm$ \tb{2.79} & 0.876 $\pm$ 0.030 & 0.755 $\pm$ \tb{0.102} & 7.218 $\pm$ \tb{1.446} & 6.938 $\pm$ \tb{0.557} & 0.914 $\pm$ 0.017 \\
		& PSDip & \tb{34.10} $\pm$ 2.83 & \tb{0.892} $\pm$ \tb{0.028} & \tb{0.760} $\pm$ 0.103 & \tb{6.744} $\pm$ 1.522 & \tb{6.275} $\pm$ 0.615 & \tb{0.940} $\pm$ \tb{0.012} \\ 
		\Xhline{1pt}
	\end{tabular}
\end{table*}

\section{Conclusion}\label{sec-conclusion}
In this work, we propose a new variational zero-shot pansharpening method PSDip. The method considers representing the HRMS by the Hamdard product of a coefficient tensor and an extended PAN. Unlike previous methods that would preset the coefficient, the proposed method regularizes the coefficient tensor in a formulated variational optimization model and optimizes it together with the expected HRMS. Specifically, a neural network takes the HRMS and the extended PAN as input and outputs the coefficient tensor, where network structure can be seen as an implicit regularisation. The formulated optimization problem is very easy to implement. After initializing the network, the network parameters and the expected HRMS are iteratively updated with regard to their respective subproblems by gradient descent methods. Experiments on benchmark datasets show the effectiveness and superiority of the proposed method.

\appendices

\section{Proof of Theorem \ref{theorem}}\label{appendix}
\begin{proof}
	As long as we find one such $R(\G)$ and $\G_2$, we will finish the proof. \\
	Let $R(\G):=\|\G - \Z\|_F^2$. The value of $\Z$ is underdetermined. Then problem \eqref{our-model} becomes
	\begin{align}\label{spe-model}
		\min_{\X,\G}~ & \left\|\Y - (\X\otimes K)\downarrow_r \right\|_F^2 + \lambda_1\left\|\X - \G\odot\hat{P}\right\|_F^2 + \nonumber \\
		&  \lambda_2 \left\|\G - \Z\right\|_F^2.
	\end{align}
	Let $\mathrm{vec}(\X)$ mean flattening a tensor $\X$ into a vector. And define
	\begin{align*}
		& \brm{y} = \mathrm{vec}(\Y), \brm{x} = \mathrm{vec}(\X), \brm{g} = \mathrm{vec}(\G), \\
		& \brm{p} = \mathrm{vec}(\hat{P}), \brm{z} = \mathrm{vec}(\Z).
	\end{align*}
	Then problem \eqref{spe-model} can be equivalently written as
	\begin{align}\label{spevec-model}
		\min_{\brm{x},\brm{g}}~\left\|\brm{y} - A\brm{x} \right\|_F^2 + \lambda_1\left\|\brm{x} - \brm{g}\odot\brm{p}\right\|_F^2 + \lambda_2 \left\|\brm{g} - \brm{z}\right\|_F^2,
	\end{align}
	where $A$ represents downsampling and bluring matrix. It is easily seen that problem \eqref{spevec-model} has an unique minimum point $(\brm{x}^*, \brm{g}^*)$. They satisfiy
	\begin{align*}
		\begin{cases}{}
			& A^T(A\brm{x}^* - \brm{y}) + \lambda_1(\brm{x}^* - \brm{g}^*\odot\brm{p}) = 0 ,\\ 
			& \lambda_1\brm{p}\odot(\brm{p}\odot\brm{g}^* - \brm{x}^*) + \lambda_2(\brm{g}^* - \brm{z}) = 0.
		\end{cases}
	\end{align*}
	Thus, we can just let 
	\begin{align*}
		 &~~~~~~~~~~\Z = \mathrm{vec}^{-1}\left(\dfrac{B}{\lambda_1\lambda_2 \brm{p}} \right), \\
		\mathrm{where}~B  = & (\lambda_1 \brm{p}^2 + \lambda_2)\odot\left[(A^TA + \lambda_1 I)\mathrm{vec}(\X^*) - A^Ty \right] \\
		& - \lambda_1^2\brm{p}^2\odot\mathrm{vec}(\X^*)
	\end{align*}
	Then problem \eqref{spe-model} has the minimum point $(\X^*, \G_2)$, where $\G_2 = \mathrm{vec}^{-1}(((A^TA + \lambda_1 I)\mathrm{vec}(\X^*) - A^Ty)/(\lambda_1\brm{p}))$.
\end{proof}

\bibliographystyle{IEEEtran}
\normalem
\bibliography{reference}

\vfill

\end{document}